\documentclass[a4paper,11pt]{article}
\pdfoutput=1 

\usepackage{jcappub} 

\usepackage[T1]{fontenc} 
\usepackage{hhline}
\usepackage{multirow}
\usepackage{amsmath}
\usepackage[table,xcdraw]{xcolor}
\usepackage{booktabs} 
\usepackage{caption}  
\usepackage{graphicx} 

\newcommand{\sevem}{\texttt{SEVEM}}

\title{\boldmath Exploring Statistical Isotropy in Planck Data Release 4: Angular Clustering and Cosmological Parameter Variations Across the Sky}

\author[a, b]{C. Gimeno-Amo,}
\author[c]{F. K. Hansen,}
\author[a]{E. Martínez-González,}
\author[a]{R. B. Barreiro,}
\author[d]{A. J. Banday}

\affiliation [a]{Instituto de Física de Cantabria, CSIC-Universidad de Cantabria,\newline Avda. de los Castros s/n, 39005 Santander, Spain.}
\affiliation[b]{Departamento de Física Moderna, Universidad de Cantabria,\newline Avda. de los Castros s/n, 39005 Santander, Spain.}
\affiliation[c]{Institute of Theoretical Astrophysics, University of Oslo, \newline PO Box 1029 Blindern, 0315 Oslo, Norway.}
\affiliation[d]{IRAP, Université de Toulouse, CNRS, CNES, UPS, Toulouse, France}

\emailAdd{gimenoc@ifca.unican.es}
\emailAdd{f.k.hansen@astro.uio.no}
\emailAdd{barreiro@ifca.unican.es}
\emailAdd{martinez@ifca.unican.es}
\emailAdd{anthony.banday@irap.omp.eu}

\abstract{The origin of small deviations from statistical isotropy in the Cosmic Microwave Background (CMB)—the so-called CMB anomalies—remains an open question in modern cosmology. In this work, we test statistical isotropy in Planck Data Release 4 (PR4) by estimating the temperature and $E$-mode power spectra across independent sky regions. We find that the directions with higher local bandpower amplitudes in intensity are clustered for multipoles between 200 and 2000 with clustering probabilities consistently below 1\% for all these scales when compared to end-to-end (E2E) Planck simulations; notably, this range extends beyond that reported in Planck Data Release 3 (PR3). On the other hand, no significant clustering is observed in the polarization $E$-modes. In a complementary analysis, we search for dipolar variations in cosmological parameters fitted using the previously computed power spectra. When combining temperature and polarization power spectra, we identify a potential anomaly in the amplitude of the primordial power spectrum, $A_{s}$, with only 5 out of 600 simulations exhibiting a dipole amplitude as large as that observed in the data. Interestingly, the dipole direction aligns closely with the known hemispherical power asymmetry, suggesting a potential link between these anomalies. All other cosmological parameters remain consistent with $\Lambda$CDM expectations. Our findings highlight the need to further investigate these anomalies and understand their nature and potential implications for better understanding of the early Universe.}

\begin{document}
\maketitle
\flushbottom

\newpage

\section{Introduction} \label{Intro}

The assumption of statistical isotropy is one of the fundamental pillars of the standard cosmological model, $\Lambda$CDM. This hypothesis is supported by the simplest inflationary models \cite{2020A&A...641A..10P}, and can be tested using observations of the Cosmic Microwave Background (CMB), as well as the large-scale structure of the Universe at late times. The analysis of CMB data from WMAP \cite{WMAP_9yr} and Planck \cite{Planck2020_I} provides a very accurate confirmation of the base $\Lambda$CDM model \cite{2020A&A...641A...6P, 2020A&A...641A...7P}, but also reveals some potential deviations from statistical isotropy in the temperature fluctuation field on large angular-scales (see \cite{2016CQGra..33r4001S, 2022JHEAp..34...49A, 2023arXiv231012859J, 2023CQGra..40i4001A} and references therein), the so called "CMB anomalies", which cannot be easily attributed to systematic effects or residual foreground contamination. Among these anomalies are the lack of correlation \cite{Hinshaw_1996, Bennett_2003, 2014MNRAS.437.2076G, 2015MNRAS.451.2978C}, also formulated as a lack of power on large scales \cite{2008MNRAS.387..209M, A.Gruppuso_2013, 2024JCAP...07..080B}, the hemispherical power asymmetry (HPA) \cite{2004ApJ...605...14E, 2004MNRAS.354..641H, Akrami:2014eta, 2016JCAP...01..046G, Marcos-Caballero:2019jqj}, the quadrupole-octopole alignment \cite{2004PhRvD..69f3516D, 2018PhRvD..98b3521M}, and a non-Gaussian feature known as the CMB Cold Spot \cite{2004ApJ...609...22V, Cruz:2004ce, 2010AdAst2010E..77V, Marcos-Caballero_2017}. Each anomaly has been tested with different estimators. Although most of the statistical tests show mild tension (2$\sigma$ - 3$\sigma$) with the $\Lambda$CDM model, some of the features seem to be uncorrelated, thus increasing the total statistical significance \cite{2016CQGra..33r4001S, 2023arXiv231012859J}. Given that intensity measurements have reached the cosmic variance limit, the largely independent information provided by large-scale polarization is  needed to clarify the origin of these anomalies. It is well established that the large angular scales in the Planck $E$-mode measurements are limited by systematic effects. Future polarization observations, such as those from LiteBIRD \cite{LiteBIRD:2020khw}, are expected to provide valuable insights into this topic.

There are three plausible explanations for the origin of these deviations. A cosmological origin is the most exciting, as it implies new physics beyond the standard model. Accordingly, there have been some attempts to explain the anomalies. For example, a cutoff in the primordial power spectrum, $\mathcal{P}$(k), could explain the lack of correlation and the low variance, while a modulation of it could produce the observed asymmetry \cite{2007ApJ...656..636G, 2020A&A...641A..10P}. Other possibilities involve multi-field inflation and the inclusion of an extra scalar field, the curvaton \cite{2008PhRvD..78l3520E, Erickcek_2009}. However, these mechanisms have their own weaknesses and none of them explain satisfactorily all of the anomalies. As an example, the non-Gaussianity and scale-invariant asymmetry predictions of curvaton models are not favored by the Planck measurements \cite{2020A&A...641A..10P} and quasar observations \cite{2009JCAP...09..011H}. A second explanation is that these anomalies could result from foreground or systematic effects. However, this is highly unlikely \cite{2016CQGra..33r4001S} as they have been observed by WMAP and Planck, despite their differing scanning strategies, systematic uncertainties, and frequency coverage. Recently, a detection of a potential new foreground \cite{2023MNRAS.518.5643L, Hansen:2023gra, Cruz:2024xbh} related to the local Universe has has been proposed as a plausible explanation for these unexpected features and could contribute to a variation of inferred cosmological parameter values over the sky \citep{toscano2025}. Other recent work \cite{2024A&A...692A.180J} has tried to quantify the impact of the Sunyaev-Zeldovich (SZ) signal from the local Universe on CMB large-scale anomalies. They conclude that the local tSZ and kSZ effects cannot account for the detected deviations from isotropy. A third possibility is that the anomalies are either statistical flukes, or that the derived significance level are not properly computed as they may be subject to a posteriori (look-elsewhere) corrections \cite{Bennett_2011}. 

In this work, we focus on another anomaly previously studied in \cite{2016A&A...594A..16P, 2020A&A...641A...7P}. Specifically, we examine the angular-clustering feature. An alignment of preferred directions derived from temperature power distribution maps has been detected across a broad range of angular scales. Following previous analyses, we compute the binned power spectrum using the \texttt{MASTER} \cite{2002ApJ...567....2H} pseudo-$C_{\ell}$ estimator in 12 independent sky regions, defined by the pixels corresponding to the $N_{\rm side}=1$ parameter of the HEALPix map scheme \cite{Gorski:2004by}. We then fit dipoles to these maps and estimate the degree of clustering between the dipole directions. We apply the same approach to polarization $E$-modes. While a dipolar distribution of power can occur in the standard cosmological model due to the Gaussian random fluctuations, the directions should be uniformly distributed. Thus, evidence for any kind of alignment between directions on different angular scales is a signature of broken statistical isotropy.

Recent studies have suggested a directional dependence of cosmological parameters \cite{2021MNRAS.504.5840F, 2022PhRvD.105h3508Y}. Given that the intensity power spectrum bandpowers appear to cluster toward the HPA direction, along with previous indications of directional variations in cosmological parameters, we test the potential presence of a dipolar feature by fitting these parameters in 12 independent sky regions. 

The outline of this paper is as follows: Section \ref{Met} describes the data, our analysis pipeline, and its validation. Section \ref{Res} presents the main results, including the angular-clustering analysis and the analysis of the cosmological parameters. In section \ref{Con}, we summarize our results and discuss their implications. Appendix \ref{Appendix_A} provides robustness tests to assess the reliability of our results.
\section{Data and Methodology} \label{Met}

\subsection{Data}

\begin{figure}[t!]
    \centering
    \includegraphics[scale = 0.85]{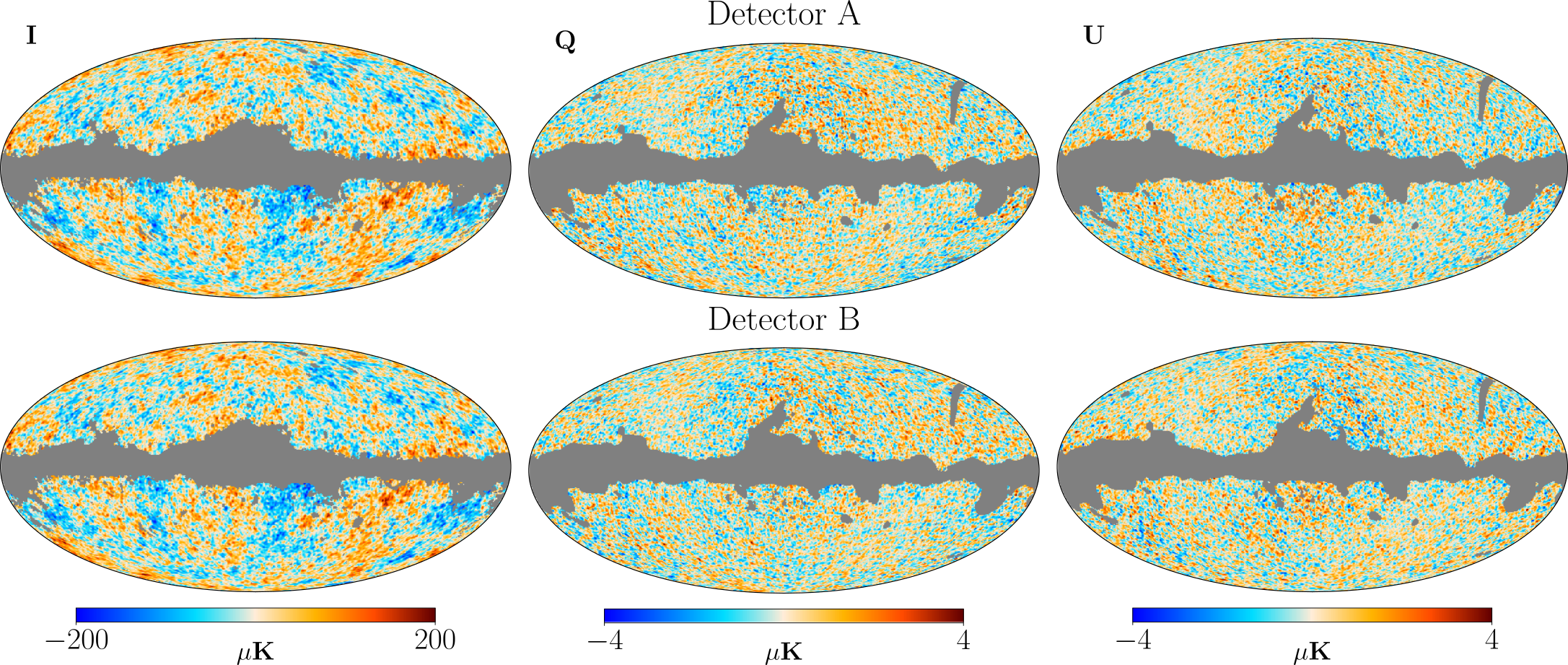}
    \caption{Planck PR4 detector A and B \sevem\ cleaned maps. First row shows the IQU maps, from left to right, for detector A, while the second row shows the B detector maps. All of the maps are smoothed with 1 degree FWHM Gaussian beam anx presented in Galactic coordinates. The grey area corresponds to the Planck PR3 confidence mask appropriate to either temperature or polarization..}
    \label{fig1}
\end{figure}

We use the data from Planck Public Release 4 (Planck PR4)\footnote{Data is available at Planck Legacy Archive (PLA), \url{https:/pla.esac.esa.int/}.}, which has been processed by the NPIPE pipeline \cite{2020A&A...643A..42P}. This new pipeline reprocesses the \textit{Planck} Low-Frequency Instrument (LFI) and High-Frequency Instrument (HFI) in a joint analysis, which effectively reduces the noise and systematics in frequency maps. In particular, we use the A and B detector splits cleaned with the \sevem\ \cite{2012MNRAS.420.2162F} component separation method. To asses the $p$-values we use the 600 available Planck PR4 "end-to-end" (E2E) simulations, which include realizations of the CMB signal, the instrumental noise, and the systematics. These simulations try to capture all the characteristics of the full data processing such as the scanning strategy and the detector responses, and sky realizations are generated also including effects such as lensing, Rayleigh scattering, and Doppler boosting. Combining independent sets of detectors allows us to use the cross-spectrum between maps avoiding noise bias. In order to mask the Galactic residual foregrounds and the extragalactic point sources, we use the PR3 Planck 2018 confidence masks described in ref.~\cite{2020A&A...641A...4P}).
These masks leave a fraction of available sky close to 78\% for both temperature and polarization. In figure~\ref{fig1} we show the detector A and B PR4 \sevem\ cleaned maps together with the Planck PR3 confidence masks.

\subsection{Analysis Pipeline}

Our pipeline can be outlined as follows:

\begin{enumerate}
    \item Power spectrum estimation:
    
    We first estimate the power spectrum of each masked map. For this purpose we use the well-known \textit{pseudo-$C_{\ell}$} \texttt{MASTER} estimator \cite{2002ApJ...567....2H}. We compute the power spectrum in bins of $\Delta{\ell}$ = 30 multipoles in the 12 equal area patches defined by the \texttt{HEALPix} \cite{Gorski:2004by} base pixels at $N_{side} = 1$. Figure~\ref{fig_pixels} enumerates each of them. The specific bin size is chosen to minimize correlations induced by the mask.
    The unmasked sky fraction, $f_{sky}$, ranges from 2\% to 8\%, depending on the overlap between the Planck confidence mask and the corresponding base pixel. Consequently, we discard the first bin $\ell$ = [2, 31]. Table~\ref{tab2} shows the exact sky fraction ($f_{\mathrm{sky}}$) values for each of the patches for both intensity and polarization. The maximum multipole is chosen to minimize contamination from point sources and to exclude multipoles where noise dominates the signal. These are our choices:

    \begin{itemize}
        \item For TT: 66 bins in the range $\ell$ = [32, 2011]
        \item For TE: 57 bins in the range $\ell$ = [32, 1741]
        \item For EE: 48 bins in the range $\ell$ = [32, 1471]
    \end{itemize}

    We applied a 0.3 degree apodization to each mask, effectively reducing correlations between multipoles at small scales in the $TT$ power spectrum. The reason behind this choice will be explained later.

    \begin{table}[t]
        \centering
        \caption{Fraction of sky ($f_{\mathrm{sky}}$) for intensity and polarization for each of the 12 considered patches.}
        \label{tab2}
        \begin{tabular}{@{}lcccccccccccc@{}}
            \toprule
            \textbf{Patch} & \textbf{0} & \textbf{1} & \textbf{2} & \textbf{3} & \textbf{4} & \textbf{5} & \textbf{6} & \textbf{7} & \textbf{8} & \textbf{9} & \textbf{10} & \textbf{11} \\
            \midrule
            $f_{\mathrm{sky}}^{T} [\%]$ & 7.2 & 7.3 & 7.8 & 7.5 & 2.3 & 4.0 & 3.4 & 4.9 & 7.6 & 7.5 & 7.3 & 7.0 \\
            $f_{\mathrm{sky}}^{P} [\%]$ & 7.7 & 7.3 & 7.7 & 7.6 & 2.4 & 4.5 & 3.5 & 5.2 & 7.8 & 7.5 & 7.4 & 7.5 \\
            \bottomrule
        \end{tabular}
    \end{table}

    \begin{figure}[t!]
        \centering
        \includegraphics[scale = 0.8]{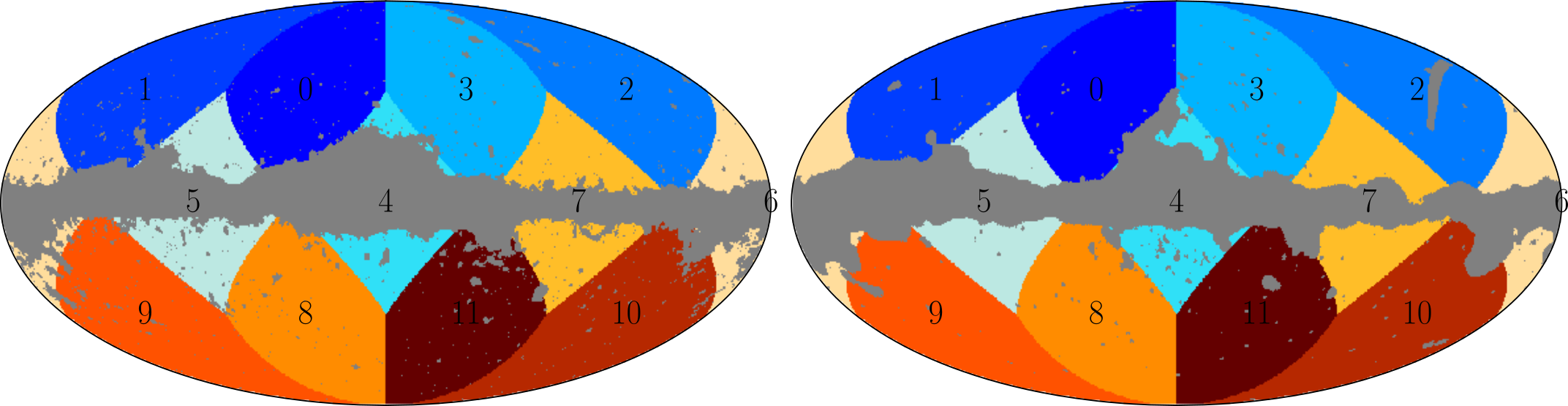}
        \caption{Numbering of the patches in a \texttt{Healpix} map at $N_{\rm{side}}$ = 1 resolution, along with the Planck confidence masks in grey. The left panel shows the intensity mask, while the right panel presents the polarization mask.}
        \label{fig_pixels}
    \end{figure}

    \item Angular Clustering: 

    We then proceed to an analysis of the Planck PR4 data to examine the angular clustering  anomaly previously noted by the Planck collaboration (see section 7.3 of \cite{2020A&A...641A...7P}). Specifically, the power spectra were computed locally in patches for various multipole ranges, and dipoles fitted to maps of the band-power estimations. Using the Rayleigh Statistic (RS), an anomalous alignment between the dipoles in the temperature data were found, with a significance level of 2–3 sigma. This behavior is not expected in the standard cosmological model, which allows for the existence of dipolar power distributions but predicts that their directions should be completely random. Therefore, this alignment is evidence of a deviation from statistical isotropy. 

    Once we have the $TT$, $TE$ and $EE$ power spectra estimated with \texttt{MASTER}, we adopt the same approach as in \cite{2020A&A...641A...7P}. 

    \begin{enumerate}
        \item For each power spectrum multipole bin, an $N_{\rm side}$ = 1 \texttt{HEALPix} map is constructed.
        \item A dipole is fitted to this map using inverse-variance weighting. The variances are computed from simulations. In particular, we are interested in the direction of the fitted dipole as the amplitudes are found to be fully consistent with the simulations.
        \item The estimator we use to measure the alignment is the modulo of the sum of all the normalized dipole vectors up to certain maximum bin,
        
        \begin{equation}\label{RSta}
            |\hat{v}| = \sqrt{N + \sum_{i \neq j} \cos{\theta_{ij}}}
        \end{equation}
        where $N$ is the number of dipoles, and $\theta_{ij}$ the angle between the $i$-th and $j$-th dipoles. This is essentially the Rayleigh statistic (RS), a statistical measure used to test uniformity, particularly for assessing whether a set of vectors exhibits any preferential alignment. Actually, eq. \ref{RSta} differs from the RS by not including any amplitude information. As previously mentioned in ref.~\cite{2020A&A...641A...7P}, the amplitude of the dipole vectors are not anomalous, so they are normalized. Apodization of the masks was essential to mitigate the clustering observed among small-scale dipoles. In simulations, where dipoles are expected to be uniformly distributed, we found that the RS value starts deviating from expectation at small scales. This deviation was later identified as artificial clustering caused by correlations between small-scale dipoles. We selected a 0.3 degree apodization scale as the minimum value necessary to recover an RS value that converges to the expected value. Minimizing the apodization scale is crucial to preserve as much of the sky as possible.
        \item Finally, we asses the clustering as a function of maximum bin using a \textit{p}-value determined as follows. We compute the RS using all the dipoles up to a certain maximum multipole bin for all the simulations and the data. Then, we define the \textit{p}-value as the fraction of simulations with a higher RS than the one observed in the data. A small \textit{p}-value means that the directions in the data are clustered in a way that can only be reproduced in a few simulations. Note that $\textit{p}$-values are correlated as they are defined from a cumulative quantity. 
    \end{enumerate}
    
    \item Parameter Estimation: 

    In order to analyze the Planck temperature and $E$-mode polarization maps to investigate possible dipolar variations of the cosmological parameters, we proceed as follows:

    \begin{enumerate}
        \item We start with the $TT$, $TE$ and $EE$ power spectra computed previously in bins of 30 multipoles in the 12 patches. The choice to use these regions for estimating cosmological parameters is guided by two primary considerations: first, the patches are disjoint eliminating correlations between them, and second, this approach is computationally efficient while remaining adequate for capturing parameter variations at the dipolar level ($\ell$ = 1).
        \item Best-fit cosmological parameters are inferred from the measured $C_{\ell}$ values in each patch. In fact, instead of adopting a traditional MCMC method, which would be computationally expensive, we use \texttt{iMinuit}\footnote{\url{https://scikit-hep.org/iminuit/}} to fit for the maximum likelihood values. We use a multivariate Gaussian likelihood, which it is a good approximation for our analysis choices ($\ell \geq$ 32),
        \begin{equation}
            -2\log{\mathcal{L}} \propto (C_{\ell}-\bar{C}_{\ell})\cdot C^{-1}_{\ell\ell'}\cdot (C_{\ell'}-\bar{C}_{\ell'})^{T}
        \end{equation}
        where $C_{\ell}$ is the observed binned power spectrum, $\bar{C}_{\ell}$ is the theoretical binned power spectrum computed from \texttt{Camb}\footnote{\url{https://camb.readthedocs.io/en/latest/}} Boltzmann solver \cite{2011ascl.soft02026L}, and $C_{\ell\ell'}^{-1}$ is the inverse of the covariance matrix. The dimension of the covariance matrix is ($n_{\mathrm{bins}}^{TT}+n_{\mathrm{bins}}^{TE}+n_{\mathrm{bins}}^{EE}$, $n_{\mathrm{bins}}^{TT}+n_{\mathrm{bins}}^{TE}+n_{\mathrm{bins}}^{EE}$). Given the size of the bins used in our analysis ($\Delta{\ell}$ = 30), we can reasonably assume that the correlations between different bins are minimal. As a result, we neglect all off-diagonal terms in all the blocks of the covariance matrix, focusing only on the variances and the covariances between the $TT$, $TE$, and $EE$ components within the same bin. 
        \item We fit for the basic flat-space $\Lambda$CDM cosmological parameters\footnote{Note that, in the likelihood, we include some bounds on the parameters to prevent the minimizer from exploring regions of the parameter space where \texttt{Camb} breaks down.}  ($H_{0}, \Omega_{c}h^{2}, \Omega_{c}h^{2}, A_{s}, n_{s}$). We fix the optical depth to reionization, $\tau$, to 0.0602, the input value for the E2E simulations, which is also in good agreement with the latest constraints \cite{Tristram:2023haj}. The main reason to fix $\tau$ is that the sky fractions for the individual patches are insufficient to assess the $E$-mode large-angular scales, where most of the information about the optical depth is encoded. We also fix $\sum m_{\nu}$ = 0.06 eV, and r = 0.01, which is also the input for the simulations. We do not expect any impact on results with this choice, as this tensor-to-scalar ratio is below the sensitivity of Planck. Additionally, following \cite{2016A&A...594A...9P} we include two effective foreground residual parameters ($A_{ps}^{TT}, A_{ps}^{EE}$), which account for residual contamination from unresolved compact objects. We assume these residuals to behave as shot-noise, modeled as $D_{\ell} \propto \ell^{2}$. 
        \item We generate an $N_{\mathrm{side}}$ = 1 map for each of the parameters, from simulations and data. We fit for a dipole in each of them using an inverse variance weighting approach, where variances are estimated from the 600 simulations. In this way, the Galactic low-latitude patches where the $f_{\mathrm{sky}}$ is smaller contribute less to the fit.
        Finally, the $p$-value is defined as the fraction of simulations with an amplitude of the fitted dipole larger or equal to the one observed in the data map.
    \end{enumerate}
    
\end{enumerate}

\subsection{Pipeline Validation} \label{sec2.3}

    In order to validate our pipeline for estimating cosmological parameters, we compare the results that we obtain using \texttt{MASTER} and \texttt{iMinuit} with those reported in \cite{2024A&A...682A..37T}. In that work, the parameters were not inferred from high-resolution foreground-cleaned CMB maps, but instead were based on likelihoods that used the cross-spectra between pairs of frequency channels. In this work, we are not interested in the absolute value of the parameters, but on their possible dipolar variation over the sky. 

\begin{figure}[t!]
        \centering
        \includegraphics[scale = 0.5]{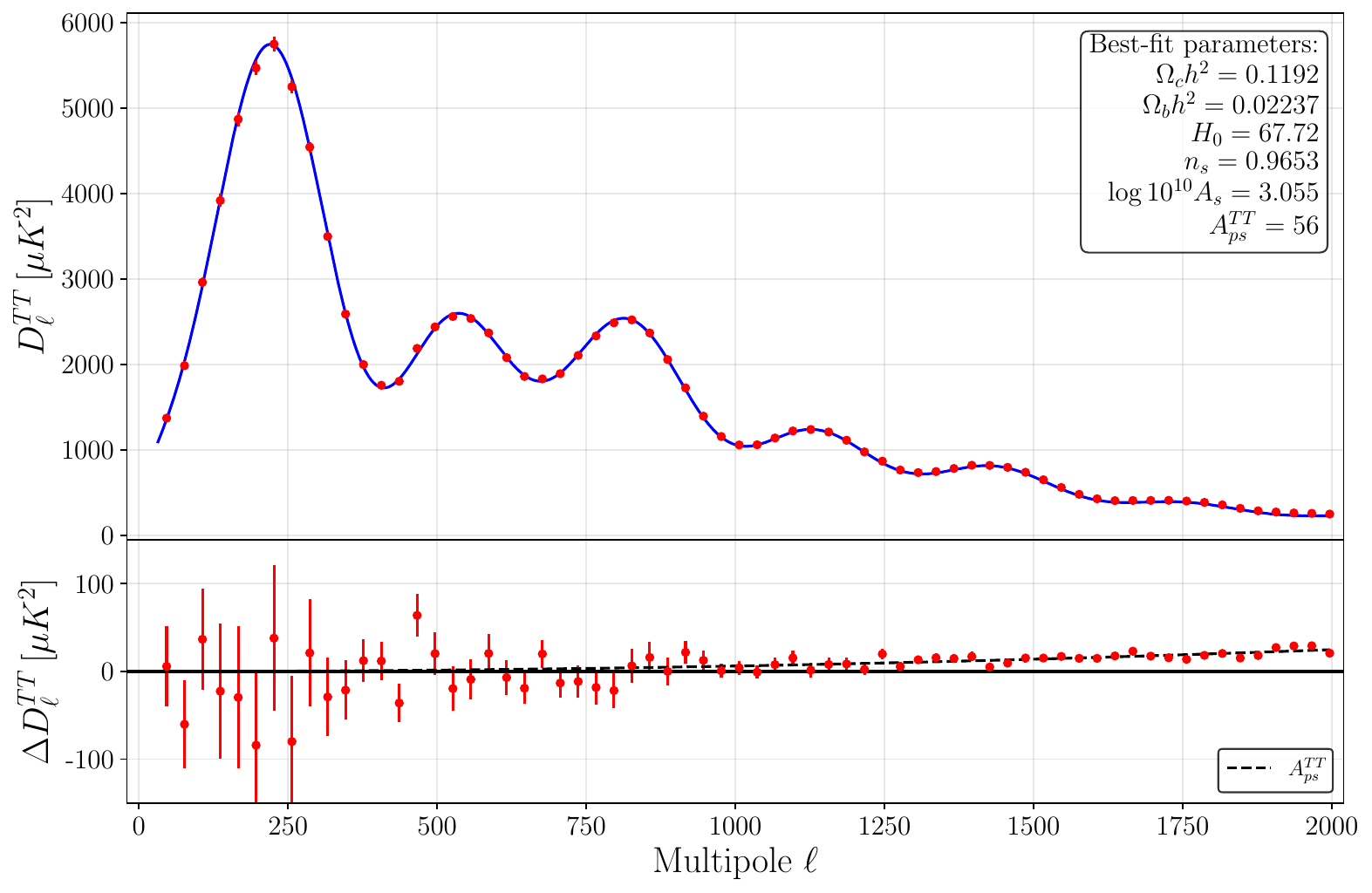}
        \caption{\textit{Upper panel}: $TT$ cross-power spectrum between A and B detectors of the PR4 \sevem\ cleaned maps (red symbols), and the best-fit $\Lambda$CDM model (blue solid line). Best-fit cosmological parameters are also provided in the box, here top right, along with the $A_{ps}^{TT}$ nuisance parameter. \textit{Lower panel}: Residuals with respect to the best-fit model. The black dashed line represents the contribution of the emission of unresolved compact objects modeled as $D_{\ell} \propto \ell^{2}$ for the fitted value of the $A_{ps}^{TT}$ nuisance parameter.}
        \label{fig2}
    \end{figure}

\begin{figure}[t!]
        \centering
        \includegraphics[scale = 0.5]{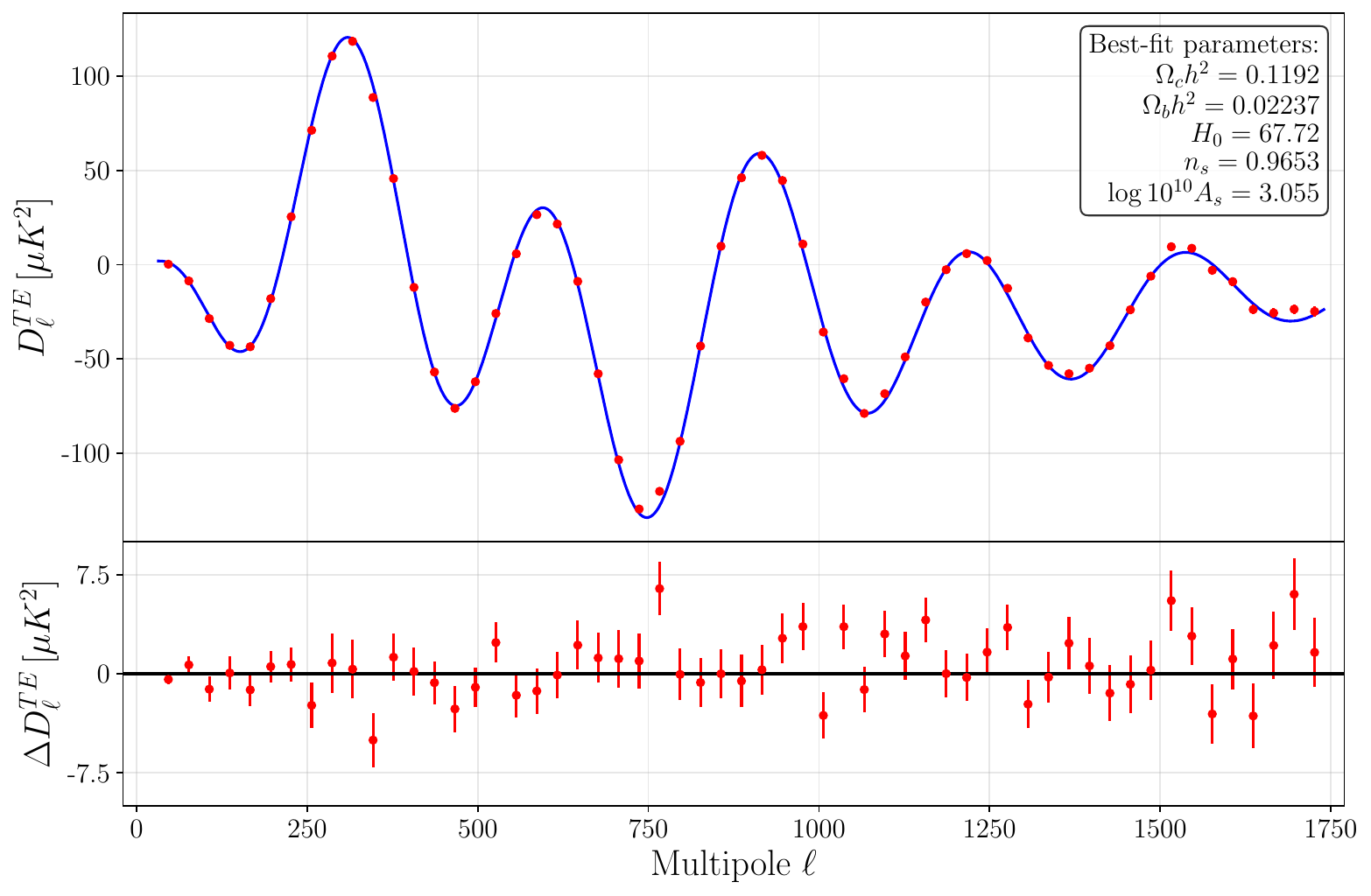}
        \caption{As figure~\ref{fig2} for $TE$.}
        \label{fig3}
    \end{figure}

\begin{figure}[t!]
        \centering
        \includegraphics[scale = 0.5]{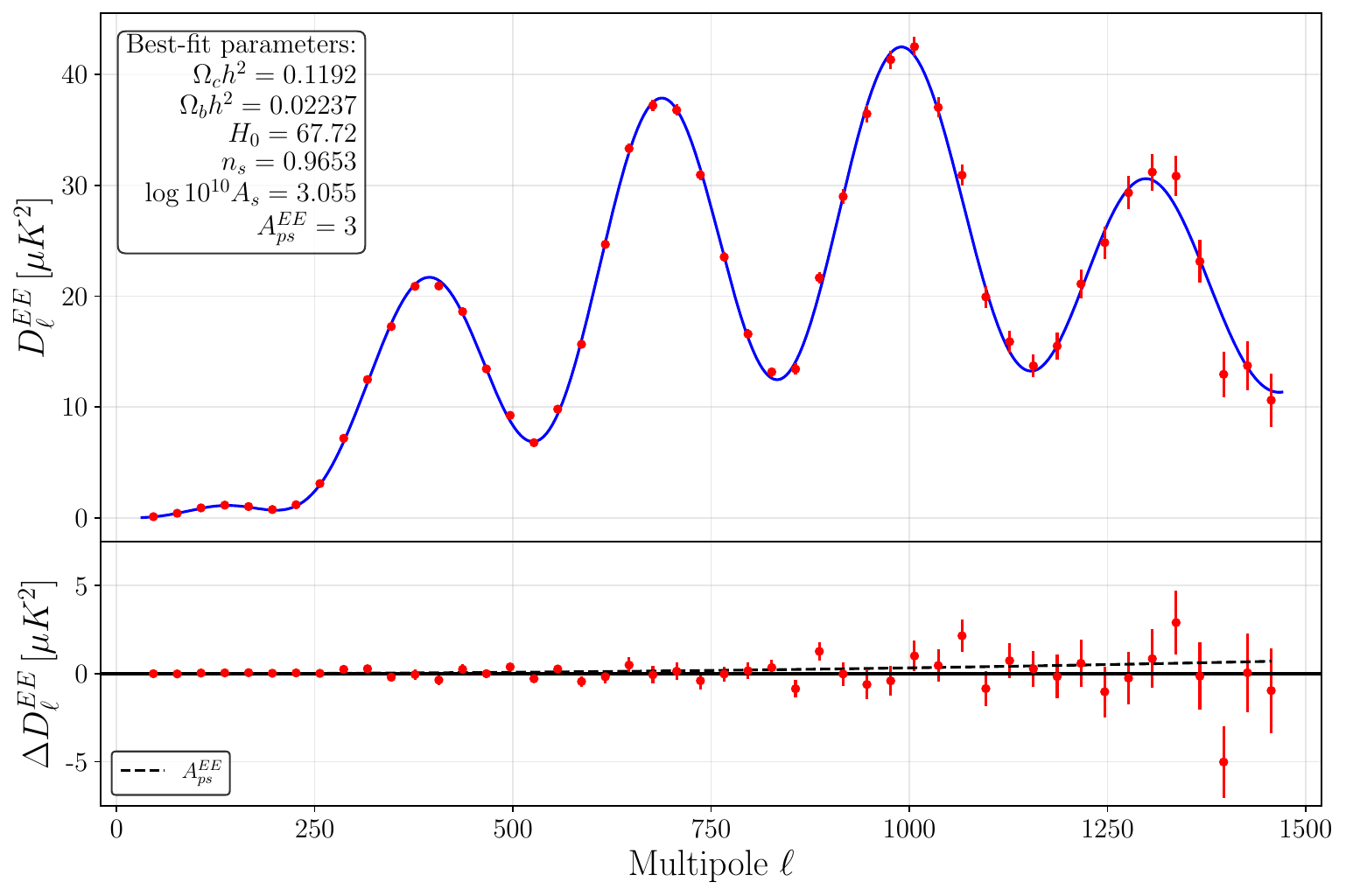}
        \caption{As figure~\ref{fig2} for $EE$.}
        \label{fig4}
    \end{figure}

    Figures~\ref{fig2}, \ref{fig3}, and \ref{fig4} show the cross-detector power spectra obtained with our pipeline for the Planck PR4 \sevem\ cleaned maps masked with the Planck confidence masks, along with the corresponding best-fit spectra computed using the cosmological parameters we obtain from \texttt{iMinuit}. The error bars are computed using the 600 E2E simulations. 

    Tests of the pipeline with simulations revealed a bias in the $TT$, $TE$, and $EE$ power spectrum estimation, which results in a small bias for the inferred cosmological parameters. We characterize this by computing the difference between the input fiducial power spectrum (based on the cosmology described in table 6 of \cite{2020A&A...643A..42P}) and the average of the output cross-spectra over the 600 E2E simulations propagated through \sevem. Subtracting this bias from each simulation, we are able to recover the input cosmological parameters. The observed bias can be explained by the mismatch between the realistic beams used in the simulations and the assumed effective beams in the analysis. Additionally, some contribution may arise from frequency-dependent effects that are not accounted for in the component separation pipeline, such as boosting. Another important point is that the bias observed in the parameters in the E2E PR4 simulations is similar to that found in the data when compared to the results of \cite{2024A&A...682A..37T}, where the analysis is performed at the frequency map level before the component separation pipeline. By this, we mean that each parameter is biased in the same direction and by a similar amount.
    The bias is not a significant concern for our analysis for two reasons. Firstly, it should affect the data and simulations in the same way. Secondly, the bias is a subdominant effect when working on small patches because the uncertainties increase given the small sky fraction. Nevertheless, we have checked the robustness of our results by fitting the parameters in the patches after debiasing the power spectra, finding consistent results (see Appendix~\ref{Appendix_A}). In this case, the bias is determined for each of the patches independently. Table~\ref{tab1} summarizes the best-fit cosmological parameters, where the error bars are estimated from the 600 E2E PR4 simulations. In particular, after correcting for the bias in the power spectrum estimation, our results (third column) are very consistent with the ones reported in \cite{2024A&A...682A..37T} (fourth column). For some parameters we are $\sim$ 1$\sigma$ away from the official Planck values, but this is expected given that we use a more limited $\ell$ range, different masks, and foreground cleaned maps. Note that the error in the $A_{s}$ parameter is a factor of 4 smaller than in \cite{2024A&A...682A..37T}. This is due to the fact that we are fixing the optical depth at reionization, which is highly correlated with the amplitude of the scalar primordial perturbations. We get similar results for the cosmological parameters when considering the weighted average over the 12 patches.

\begin{table}[t]
\centering
\caption{Best-fit cosmological parameters. The second and third columns present the results from our pipeline with and without bias correction (no debiasing and debiased, respectively). These values are obtained using the Planck PR3 confidence masks. The fourth column shows the latest cosmological parameters derived in ref.~\cite{2024A&A...682A..37T} from the $TTTEEE$ power spectra using the PR4 dataset. The last two rows provide the results for the nuisance parameters which are given by their values at $\ell = 3000$. The error bar in $A_{s}$ for the first and second columns are much smaller because $\tau$ is fixed.}
\label{tab1}
\begin{tabular}{|l|ccc|}

\hline
Parameter         & TTTEEE (no debiasing) & TTTEEE (debiased) & PR4 (TTTEEE)    \\ \hline
$H_{0}$           & $66.78\pm0.50$       & $67.72\pm 0.50$     & $67.64\pm 0.52$      \\
$\Omega_{b}h^{2}$ & $0.02212\pm0.00013$  & $0.02237\pm 0.00013$   & $0.02226\pm 0.00013$ \\
$\Omega_{c}h^{2}$ & $0.1209\pm0.0011$    & $0.1192\pm 0.0011$    & $0.1188\pm 0.0012$   \\
$\ln{(A_{s}\cdot 10^{10}})$           & $3.057\pm0.0033$    & $3.055\pm 0.0033$    & $3.040\pm 0.014$     \\
$n_{s}$           & $0.9598\pm0.0036$    & $0.9653\pm 0.0037$    & $0.9681\pm 0.0039$   \\
$A_{ps}^{TT}$     & $55\pm4$           & $56\pm 4$       & -                    \\
$A_{ps}^{EE}$     & $0\pm1$            & $3\pm 1$        & -                    \\ \hline
\end{tabular}

\end{table}

\subsection{Bayesian Approach}

\begin{figure}[t!]
    \centering
    \includegraphics[scale = 0.43]{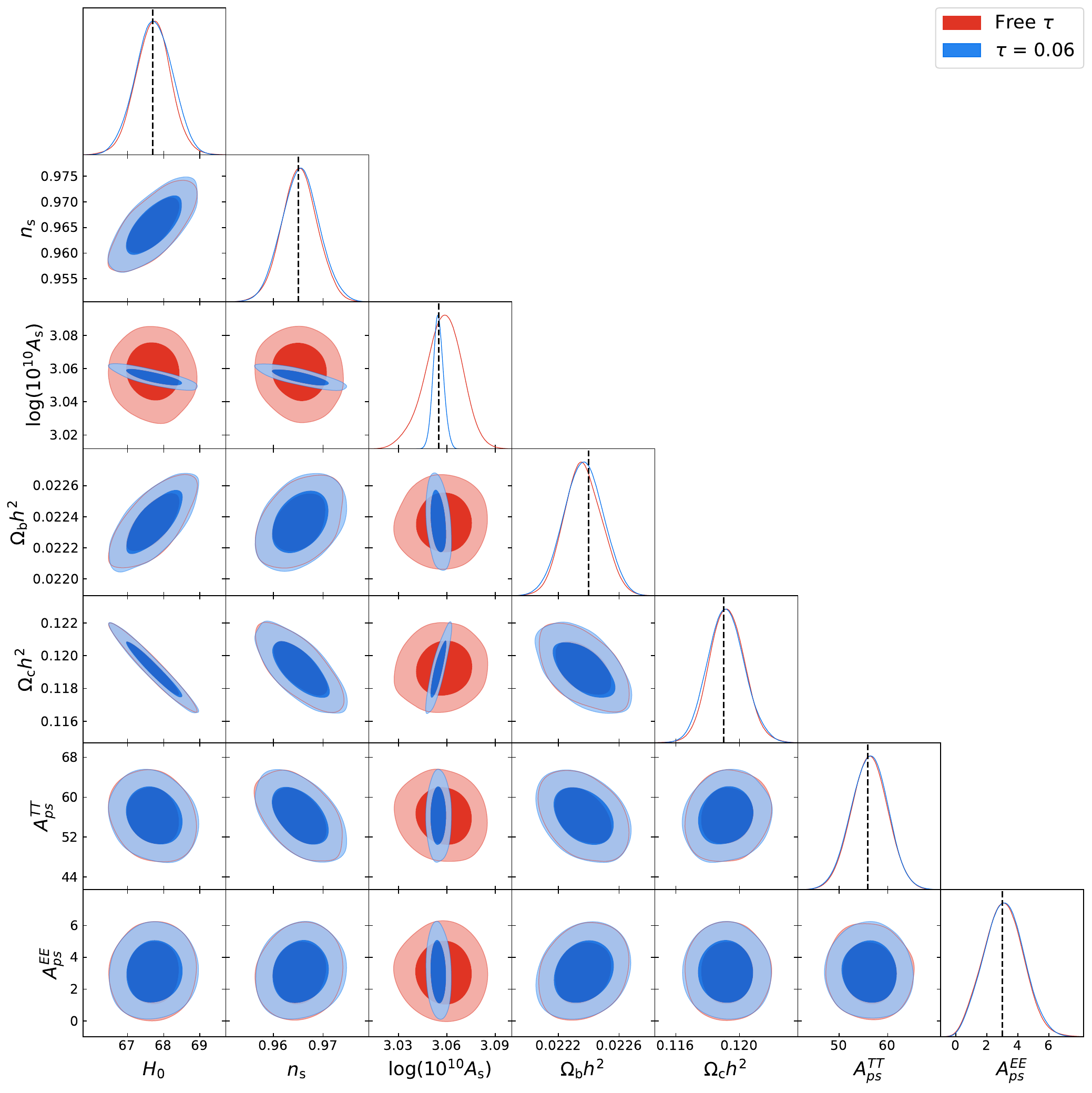}
    \caption{Constraints on the base $\Lambda$CDM model using the $TT$, $EE$, and $TE$ cross-spectra between detector A and B for the \sevem\ PR4 cleaned maps. A binned power spectrum, starting from $\ell$ = 32, and a Gaussian likelihood are used. Diagonal plots are the marginalized parameter constraints. Black dashed lines correspond to the parameters obtained with \texttt{iMinuit}. Two scenarios are considered: fixing $\tau=0.06$ (in red) and leaving $\tau$ as a free parameter (in blue). Contours contain 68\% and 95\% of the probability.}
    \label{fig5}
\end{figure}

We perform an MCMC analysis on the debiased data as an extra validation of our pipeline. For this purpose, we use \texttt{cobaya}\footnote{\url{https://cobaya.readthedocs.io/en/latest/}} \cite{2002PhRvD..66j3511L, 2013PhRvD..87j3529L, 2021JCAP...05..057T}. Figure~\ref{fig5} shows the posteriors for each of the parameters together with the value inferred from \texttt{iMinuit}. We use flat priors in the same region where minimization is performed, and compute the posteriors for two scenarios: fixing $\tau$ (red contours), and leaving it free but with a Gaussian prior $\mathcal{N}(0.06, 0.006)$ (blue contours) applied. This test reveals three key conclusions. First, the minimum found by \texttt{iMinuit} is fully consistent with the position of the peak in the posterior for all parameters, showing the robustness of the minimizer. Second, the width of the posterior aligns remarkably well with the standard deviation of the minimum values obtained from the 600 simulations. Finally, the posterior width for $A_{s}$ is significantly reduced when $\tau$ is fixed.

\section{Results} \label{Res}

\subsection{Angular-Clustering}

\begin{figure}[t!]
    \centering
    \includegraphics[scale = 0.8]{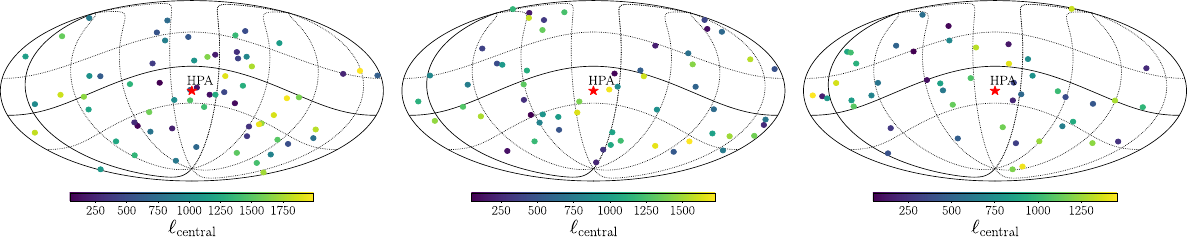}
    \caption{Dipole directions for $\Delta\ell$ = 30 bins of the distribution of the cross power spectrum between detector A and B splits of the \sevem\ PR4 cleaned maps, from $\ell$ = 2 to $\ell$ = 2011 ($TT$), 1741 ($TE$), and 1471 ($EE$). The direction for a specific multipole bin is coloured according to the central value of the bin, as shown in the colour bar. The maps are rotated in such a way that the center is located in ($l$, $b$) = (205, -20) in Galactic coordinates, which is the preferred direction for the HPA in temperature data (marked with a red star). The left panel shows the directions for $TT$, the middle panel for $TE$, and the right panel for $EE$. Graticule shows the Galactic reference frame.}
    \label{fig7}
\end{figure}

\begin{figure}[t!]
    \centering
    \includegraphics[scale = 0.4]{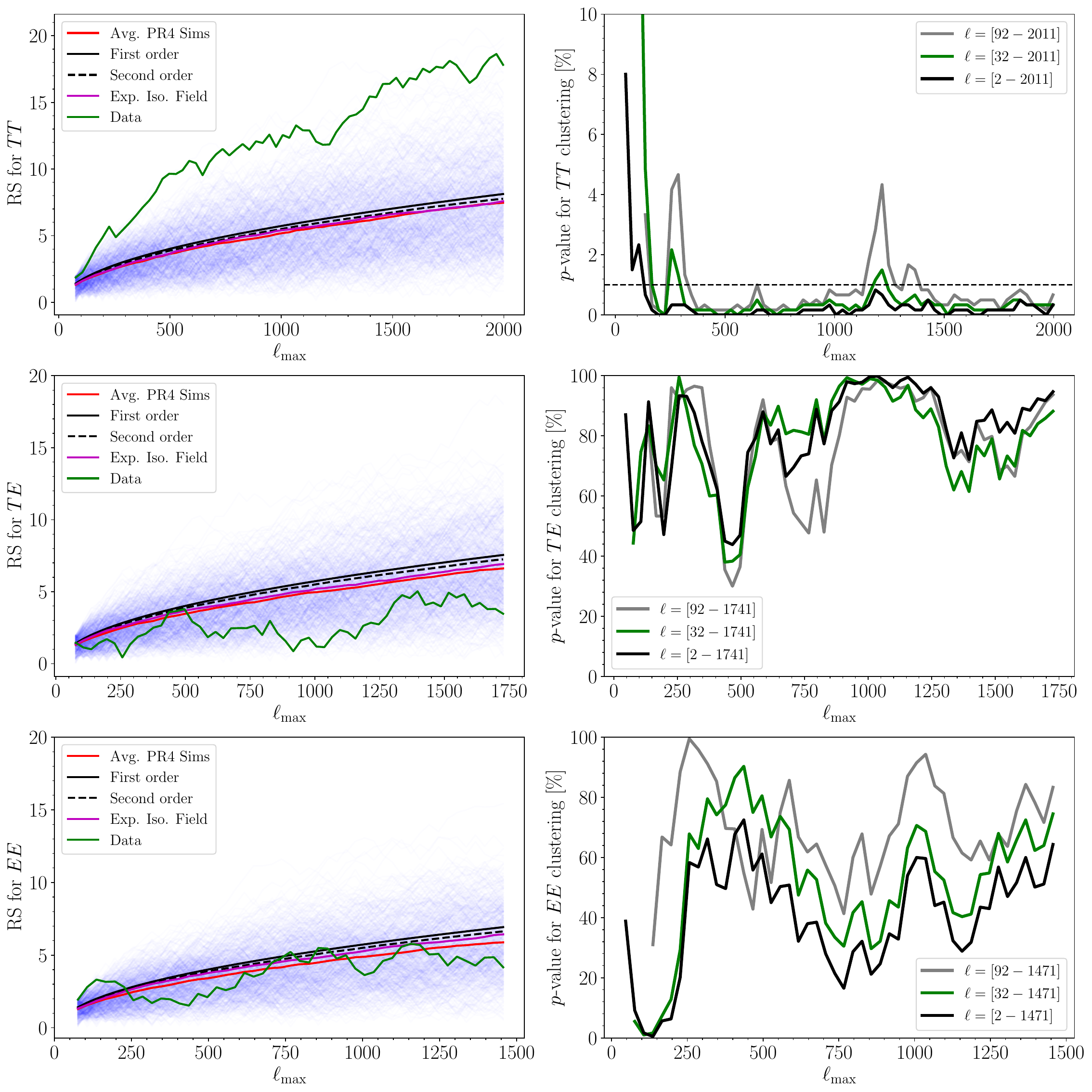}
    \caption{\textit{Left column}: RS estimator for $TT$ (top panel), $TE$ (middle), and $EE$ (lower). Blue curves correspond to each of the 600 E2E PR4 simulations, while the red line is their average. Results for the data are shown by the green line. The purple curve is the expected value for an isotropic field. The solid (dashed) black line is the theoretical expectation at first (second) order.
    \textit{Right column}: Derived $p$-values for the angular-clustering of the cross-power distribution obtained from the PR4 detector split maps as a function of $\ell_{\mathrm{max}}$. The $p$-values are derived from the fraction of E2E PR4 simulations with an RS equal to or larger than that observed in the data, hence small $p$-values would correspond to an unexpected alignment between dipole directions. The black line shows the results starting from $\ell$ = 2, while the green and grey curves start in $\ell$ = 32 and $\ell$ = 92, respectively. For the $TT$ case, the black dashed line represents the 1\% $p$-value.}
    \label{fig6}
\end{figure}

In this section, we show the results for the angular-clustering analysis. The three panels in figure~\ref{fig7} 
show the $TT$, $TE$, and $EE$ dipole directions determined in bins of $\Delta\ell$ = 30 for the A/B detector splits of the \sevem\ PR4 data. The plots are rotated in such a way that the center of the image is located at ($\ell$, b) = (205, -20) in Galactic coordinates. This is the direction of the HPA found in the temperature data in ref.~\cite{2023JCAP...12..029G}. The left column of figure~\ref{fig6} presents the corresponding RS values as a function of $\ell_{\mathrm{max}}$, while the right column shows the associated $p$-values. We consider as our reference case that obtained without including the first bin (2 $\leq$ $\ell$ < 32), which is given by the green line in the left and right panels. In the same figure, we include the expected RS curve for an isotropic field, which has been computed from random directions uniformly distributed in the sky. Additionally, the theoretical expectation, at first and second order, for the average value across the simulations is included.
\begin{equation}
    \left<\sqrt{N+2x}\right> \approx \sqrt{N} - \frac{1}{4N^{3/2}}\sigma_{x}^{2} + \mathcal{O}(\left<x^{3}\right>)
\end{equation}
where x = $\sum_{ij}\cos{\theta_{ij}}$, and $\sigma_{x}^{2}$ is the variance of x. We are also assuming $<\sum_{ij}\cos\theta_{ij}> = 0$. 

\begin{figure}[t!]
    \centering
    \includegraphics[scale = 0.6]{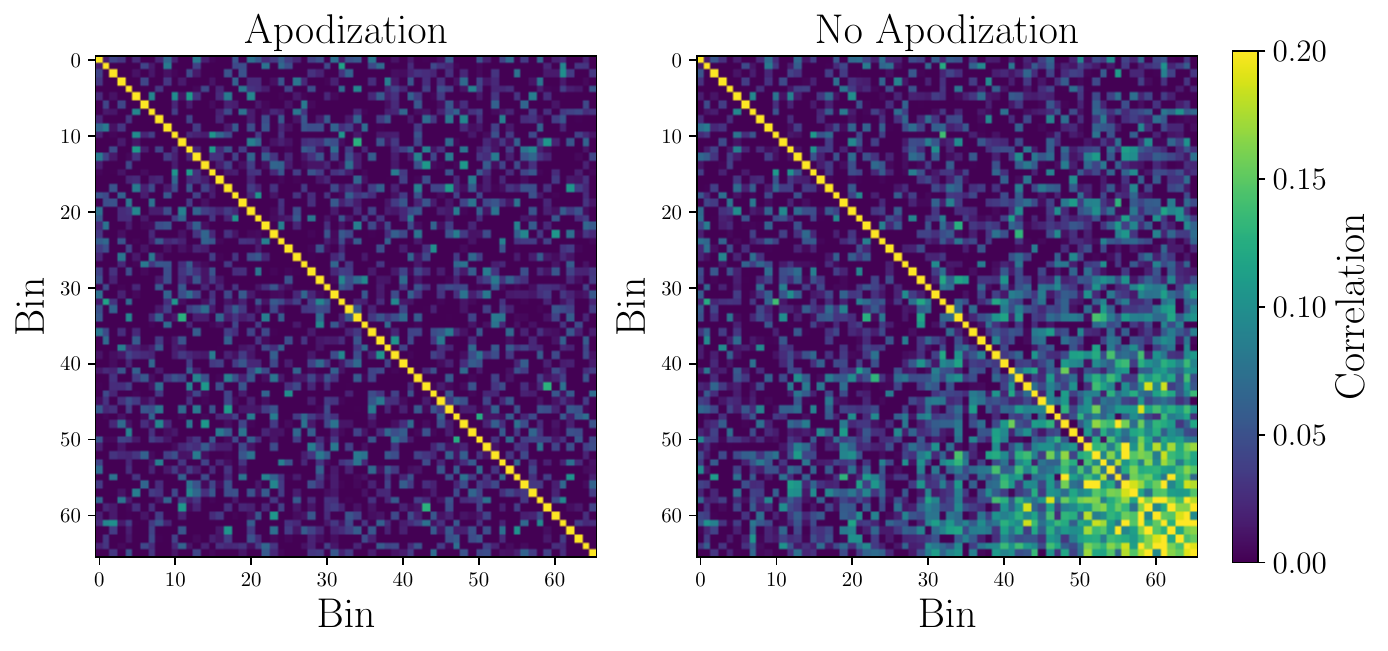}
    \caption{Correlation matrix between bins of $\Delta\ell$ = 30 for the temperature power spectrum. The left panel shows the results for the apodization case (0.3$^\circ$), while the right panel corresponds to the no apodization scenario.}  
    \label{fig8}
\end{figure}

The significance of the temperature alignment is compatible with previous results up to $\ell_{\rm{max}} \approx$ 750. From $\ell_{\mathrm{max}} \approx$ 200 the $p$-value is essentially below 1\% up to the maximum multipole considered. We also consider two additional cases in which we slightly modify the first bin included in the analysis. The black line in the left panels corresponds to the case where we include the first bin, thus incorporating all the information from $\ell$ = 2 to 2011. In contrast, the grey line represents the case where we exclude the first three bins. In both cases, the results appear to be robust, with the $p$-value remaining below 1\% for most multipoles. Only two regions, around $\ell \approx$ 300 and $\ell \approx $ 1200, show a slight increase in the $p$-value. Since the $p$-value is a cumulative quantity, this could be due to the dipoles in these bins being oriented far from the clustering direction, effectively contributing negatively to the RS. In a previous Planck paper \cite{2020A&A...641A...7P}, the $p$-value was found to increase rapidly from $\ell_{\mathrm{max}} \approx$ 1000. We observe a similar increase in the case where we do not apodize the mask. Figure~\ref{fig8} presents the correlation between the bins in temperature for both scenarios, with and without apodization. For small scales, we see an increase of correlations for the non-apodized case, which produces an artificial clustering in simulations, and thus reduces the significance. Additionally, we note that for $\ell_{\mathrm{max}}$ < 100 the temperature $p$-values are not anomalous, which is inconsistent with the HPA reported in the analysis of large angular scales. As mentioned in ref.~\cite{2020A&A...641A...7P}, this could simply be due to the high variance of the estimator in this region.

An important step during the dipole fitting is the subtraction of the mean field for each bandpower. We realized that the mean fields exhibit a dipolar feature pointing in the direction of the CMB dipole. This could be due to the Doppler boosting effect \cite{2016A&A...594A..12P, 2002PhRvD..65j3001C}, which is also simulated in the FFP10 realizations. We checked the impact of the mean field subtraction in the dipole directions. anf found that the distribution of dipoles indeed exhibits significant non-uniformity when the mean field is not subtracted during the fitting. However, this is corrected after the mean field subtraction.

It is evident from the right panel of figure~\ref{fig6} that some $p$-values for the $TE$ spectrum are close to 100\%. This observation was also reported in the previous Planck analysis. A high $p$-value means a low value for the RS statistic, which could also be anomalous. To assess whether such low values are unusual, we examine the maximum $p$-value of the RS statistic for the $TE$ data, which is 99.5\%, and scan over the simulations to see in how many of them we are able to find such high $p$-values\footnote{Note that the $p$-value of the simulations is computed by removing the given simulation from the set and using the remaining 599 as the reference.}, but not restricting ourselves to any $\ell_{max}$ range (to take into account the look-elsewhere effects). In fact, we find that the maximum $p$-value in the data is exceeded in 16\% of the simulations. In other words, 16\% of simulations show at least one $p$-value above 99.5\%. Furthermore, we observe in the data that in the multipole range $\ell_{\mathrm{max}}$ between 900 and 1050, $p$-values consistently fall above 95\%. We analyze the simulations finding that approximately 6\% of them exhibit a range of 7 consecutive bins with $p$-values above the 95\% threshold. These results indicate that neither of these features is statistically anomalous, and can be explained by look-elsewhere effects.

For the $EE$ polarization signal, the $p$-value is at the 1\% level only for a single bin ($\ell_{\mathrm{max}}$ = 62-91). For the case where the first bin is included the $p$-value reaches a minimum $p$-value of 0.5\% again for a single bin ($\ell_{max}$ = 92-121). However, we do not see any anomalous behavior in that range for either $TT$ or $TE$. Considering that in PR3 the $p$-value in the $E$-modes remained close to 1\% for several consecutive bins, whereas here it is observed in a single bin, which is not statistically significant, most likely the PR3 results were more affected by systematics at low-$\ell$ in polarization. We conclude that the hint of an anomaly in the $E$ modes observed in PR3 has disappeared in the PR4 data.

Following the analysis performed ref.~\cite{2020A&A...641A...7P}, we also tested whether the directions of the $EE$ dipoles are aligned with the $TT$ dipoles. Here we made a small change in the statistic described in section~\ref{Met}. In order to avoid using the information from $TT$ and $EE$ alone, we simply use the mean of the cosine of the angles between all pairs of dipoles, where one is $TT$ and the other one is $EE$, i.e. $\left<\sum_{ij} \cos{\theta_{ij}}\right>$, where $\theta_{ij} = v_{i}^{TT} \cdot v_{j}^{EE}$.

\begin{figure}[t!]
    \centering
    \includegraphics[scale = 0.5]{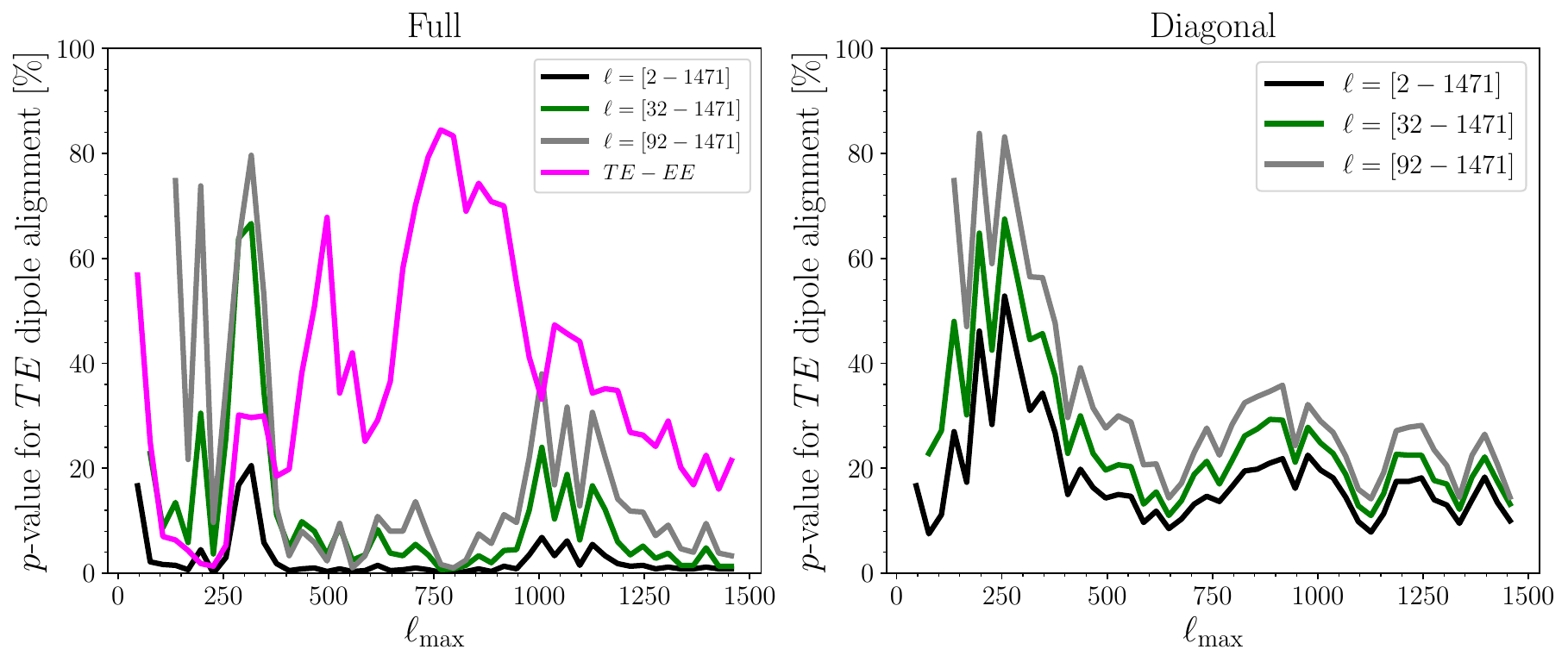}
    \caption{Derived $p$-values for the angular-clustering between $TT$ and $EE$ dipole directions. Black, green, and grey colors corresponds to the same cases as in Figure \ref{fig8}. The magenta line shows the corresponding correlation between the $TE$ and $EE$ dipole directions in the $\ell$ = [2-1471] range. The $p$-values are derived from the fraction of E2E PR4 simulations with a RS value equal to or larger than the one observed in the data. Results in left panel are obtained by averaging the cosines between all the dipole directions, while for right panel only cosines between dipoles within the same bin are used.}
    \label{fig9}
\end{figure}

Figure~\ref{fig9} shows the $p$-values for $TT$-$EE$ alignment. The main motivation for studying such alignment in previous Planck works was the existence of a multipole range below $\ell_{\rm{max}}$ = 250 where the $p$-value for both $TT$ and $EE$ was below 1\%. In this work, such a situation does not occur, but we still find it interesting to perform the same analysis.

The black curve in figure~\ref{fig9}, corresponding to the including the first bin, exhibits a pattern similar to that shown in figure~40 of \cite{2020A&A...641A...7P} up to $\ell_{\mathrm{max}} \approx $ 1000. This indicates that $TT$ and $EE$ appear to be clustered towards a similar direction at the level of 1\% over a wide range of $\ell_{\rm{max}}$. This is highly unexpected if $TT$ and $EE$ are completely independent, even if they were clustered individually. However, we know that a non-zero $TE$ spectrum induces a correlation between $T$ and $E$. In the Planck paper, this was explored with simulations to assess whether the $TT$-$EE$ alignment is expected in the case where both $TT$ and $EE$ are individually clustered. In particular, they examine all simulations having a minimum $p$-value below 1\% for both $TT$ and $EE$ in overlapping multipole ranges, and then check the correlation between $TT$ and $EE$ directions. They just found 2 simulations satisfying that criterion, with neither showing a high correlation between directions. Furthermore, we also explore the $TE$-$EE$ alignment (magenta line in figure~\ref{fig9}), and do not find anomalous features. The results seem to lose statistical significance once the first bins are removed (see green and grey lines in figure~\ref{fig9}).

Another interesting result is shown in the right panel of figure~\ref{fig9}. In this case, only the cosines between the $TT$ and $EE$ vectors within the same bin are considered, that is, when i = j. No anomalous alignment behavior is observed, as the $p$-value never drops below 10\%. This suggests that the anomaly mainly arises due to the off-diagonal terms.

Our results show a clear anomaly in the $TT$ clustering. On the other hand, $TE$ and $EE$ seem to be compatible with the Planck E2E PR4 simulations, and only exhibit anomalous behavior over a very narrow multipole range, which suggests a look-elsewhere effect. Similarly, the $TT$-$EE$ alignment indicates, under certain conditions, a clear anomaly with $p$-values below 1\%. However, the interpretation is not entirely clear, as the statistical significance seems to depend slightly on whether the first bins are included or not. Moreover, if only the cosines between dipoles within the same bin are considered, the statistical significance is greatly reduced.

\subsection{Analysis on Cosmological Parameters}

\begin{figure}[t!]
    \centering
    \includegraphics[scale = 0.5]{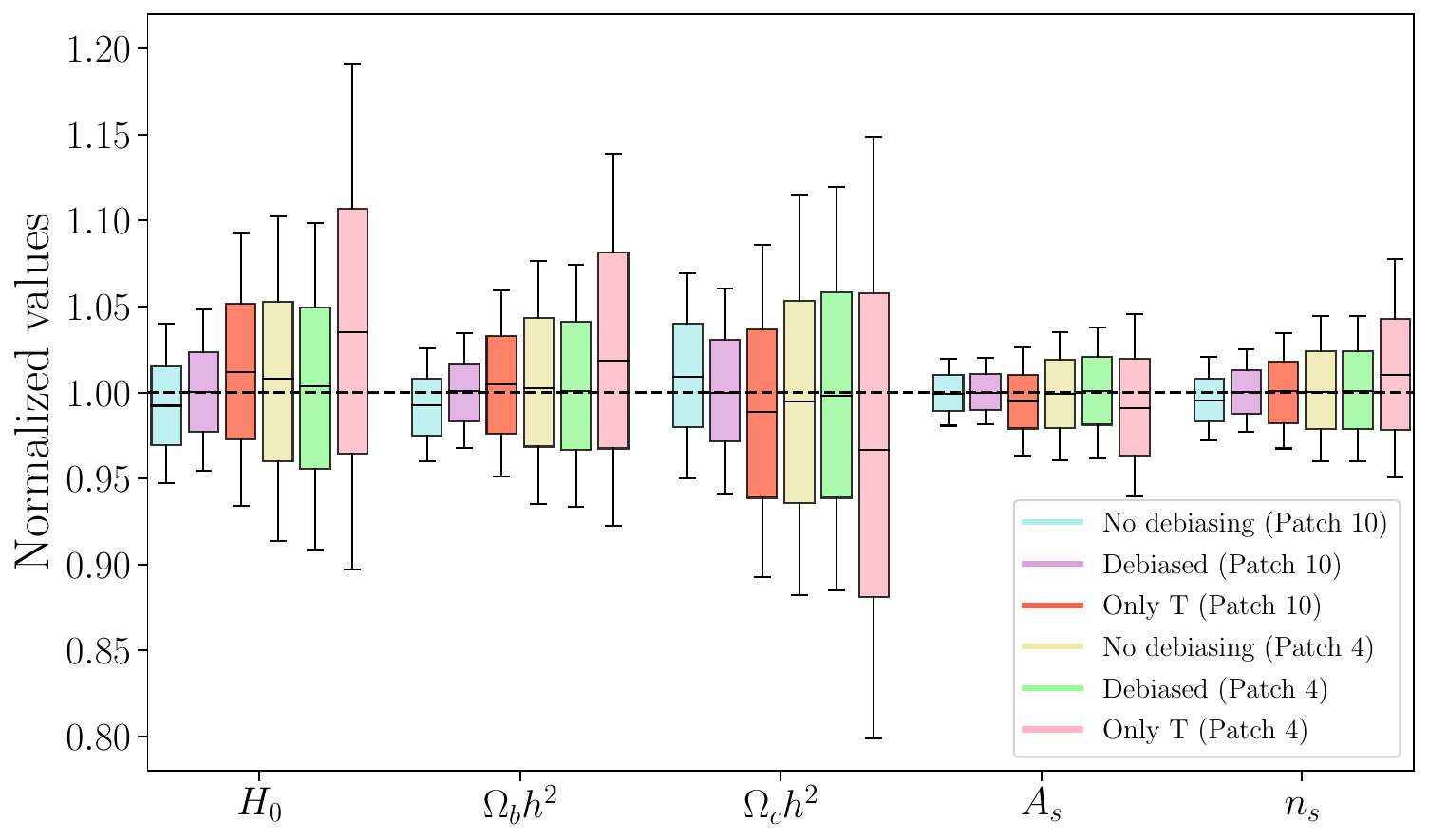}
    \caption{Distribution of the cosmological parameters for the 600 E2E PR4 simulations for two different patches, patch 4 ($f_{\mathrm{sky}} \approx $ 2\%) and patch 10 ($f_{\mathrm{sky}} \approx $ 7\%). For comparison three different scenarios are included: temperature only (no debiasing), including polarization (no debiasing), and the debiased case (corrected for the bias in temperature and polarization). The distributions are normalized to the input values. The boxes represent 68\% of the probability, while the large error bars include 95.4\%.}
    \label{fig10}
\end{figure}

\begin{figure}[t!]
    \centering
    \includegraphics[scale = 0.5]{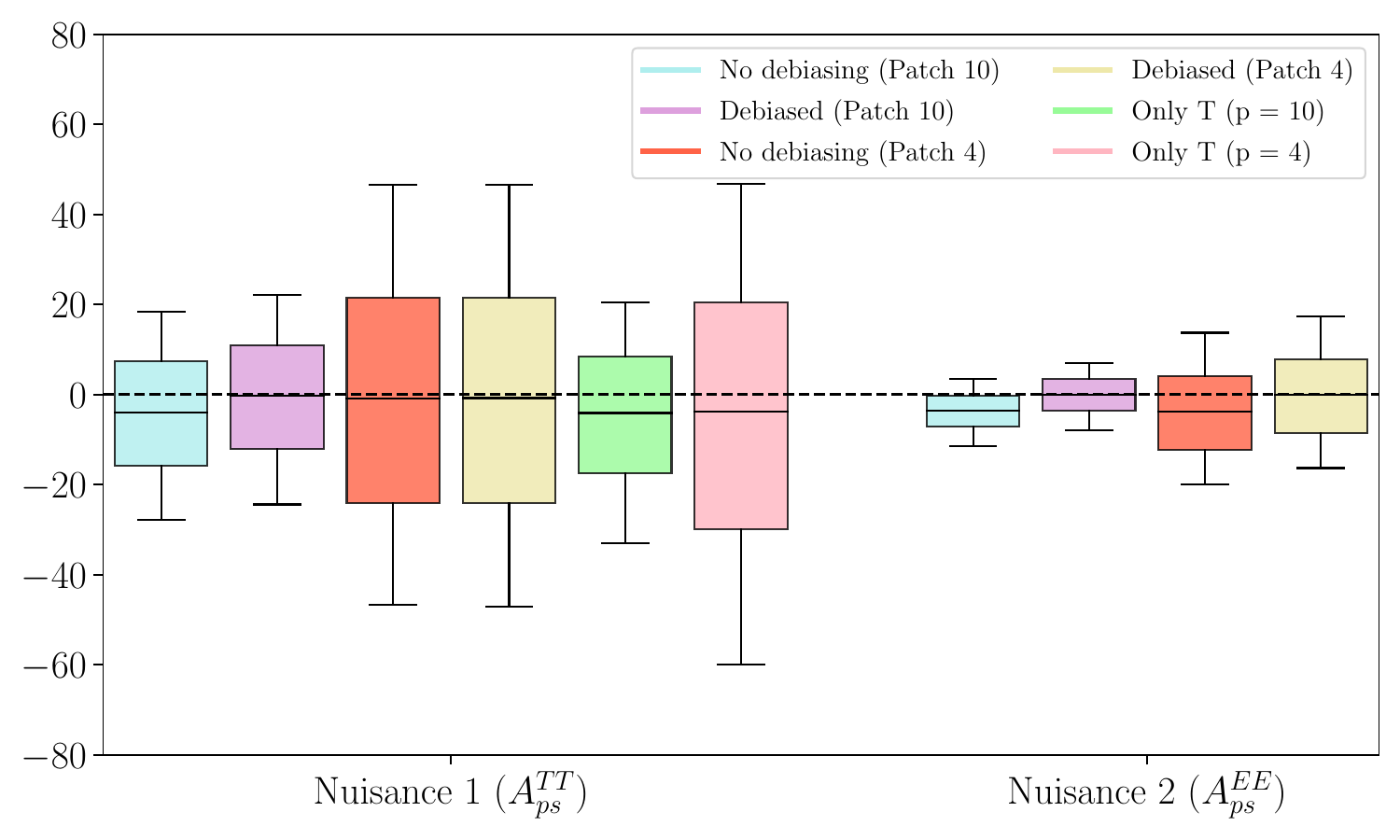}
    \caption{Same as figure~\ref{fig10} for the nuisance parameters.}
    \label{fig11}
\end{figure}

\begin{figure}[t!]
    \centering
    \includegraphics[scale = 0.8]{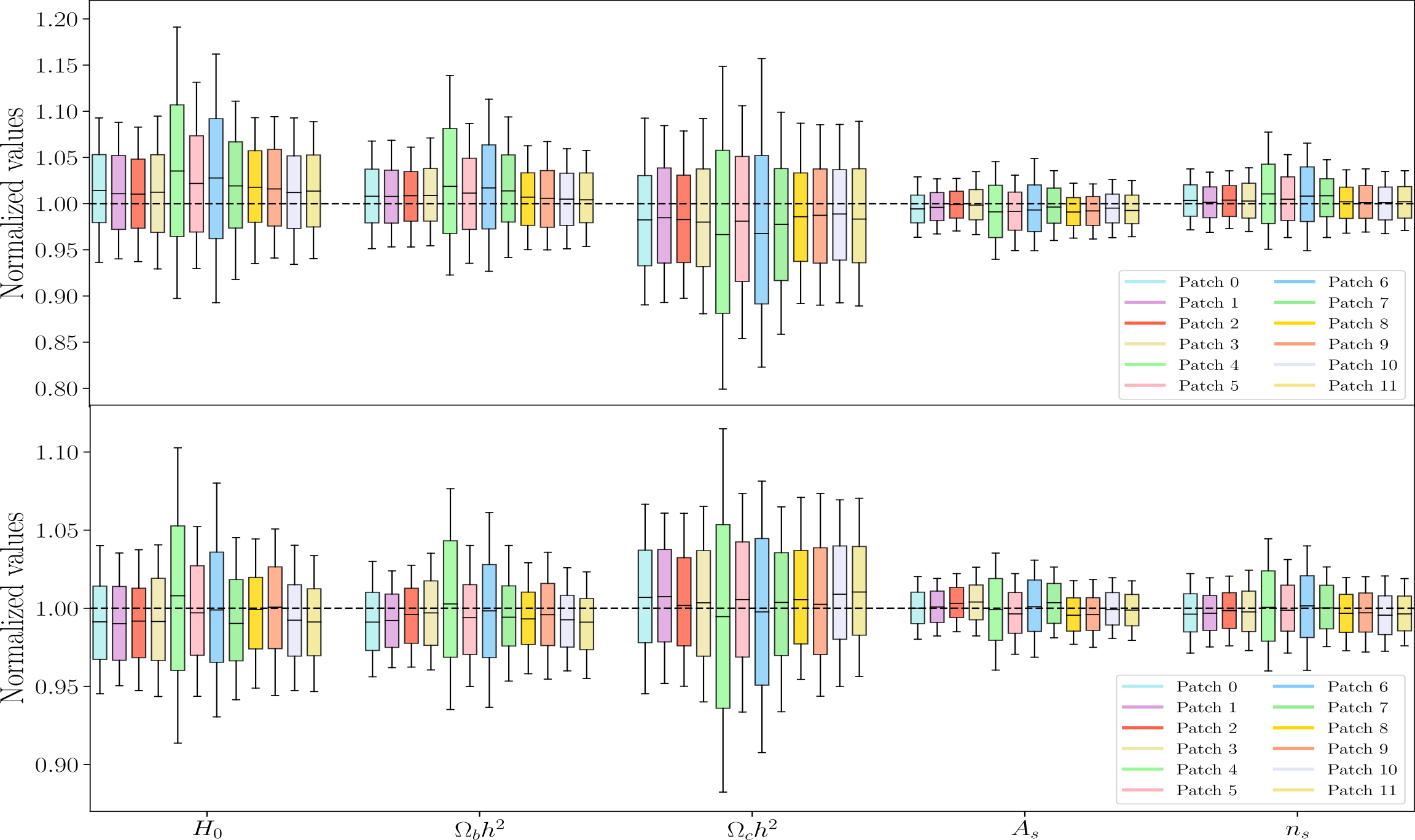}
    \caption{Distribution of the cosmological parameters computed from 600 E2E PR4 simulations for the 12 patches in the temperature only case (top panel) and including polarization (bottom panel). In both cases, no-debiasing results are shown. The distributions are normalized to the input values. The boxes represent 68\% of the probability, while the large error bars include 95.4\%.}
    \label{fig15}
\end{figure}

In this section, we show the results for the analysis of the cosmological parameters. We will consider as reference the case where we do not subtract the bias (see section~\ref{sec2.3}) at the power spectra level. Figures~\ref{fig10} and \ref{fig11} show the cosmological and nuisance parameters, respectively, for patch 4 and 10. These plots provide, for the patch with the smallest sky coverage and one of those with the largest, a direct comparison between the debiased and no debiasing cases. For reference, we also include the only temperature scenario, which has been run without including the $TE$ and $EE$ power spectra. As mentioned in section~\ref{sec2.3}, the bias on the power spectra could have a significant impact on the cosmological parameters for large $f_{\mathrm{sky}}$. However, for smaller sky fractions, the bias is subdominant compared to the statistical uncertainties. In particular, figure~\ref{fig15} shows the distribution of the 5 cosmological parameters for all patches for the two no-debiasing cases: only temperature (top panel) and including polarization (bottom panel). The impact of the bias on the parameters can be directly observed in these plots. The parameters are not strongly affected compared to the large error bars. Note also that most of the patches are biased in a similar way, so the bias is not expected to produce a dipolar pattern. Nevertheless, such a pattern would be removed in the dipole fitting process by subtracting the mean field. This is the main reason why the results are robust against the debiasing (see appendix~\ref{Appendix_A}). Note also the dependence of the error bars with the patch index and sky fraction (see table~\ref{tab2} and figure~\ref{fig_pixels}). It is clear from figure~\ref{fig10} that the bias only produces a shift of the distribution while, as expected, a lower sky fraction leads to a broadening of it. In appendix~\ref{Appendix_A} we perform a set of robustness tests by running the pipeline for different analysis choices, including the debiased cases and cuts in the $\ell_{min}$ and $\ell_{max}$, showing that the results are in general quite stable. Note in Figure \ref{fig11} that, in the no debiasing case, the nuisance parameters tend to have a negative average value. Since point sources have not been simulated, these parameters should be zero. This reflects how they attempt to absorb the effect of the bias.

\begin{figure}[t!]
    \centering
    \includegraphics[scale = 0.9]{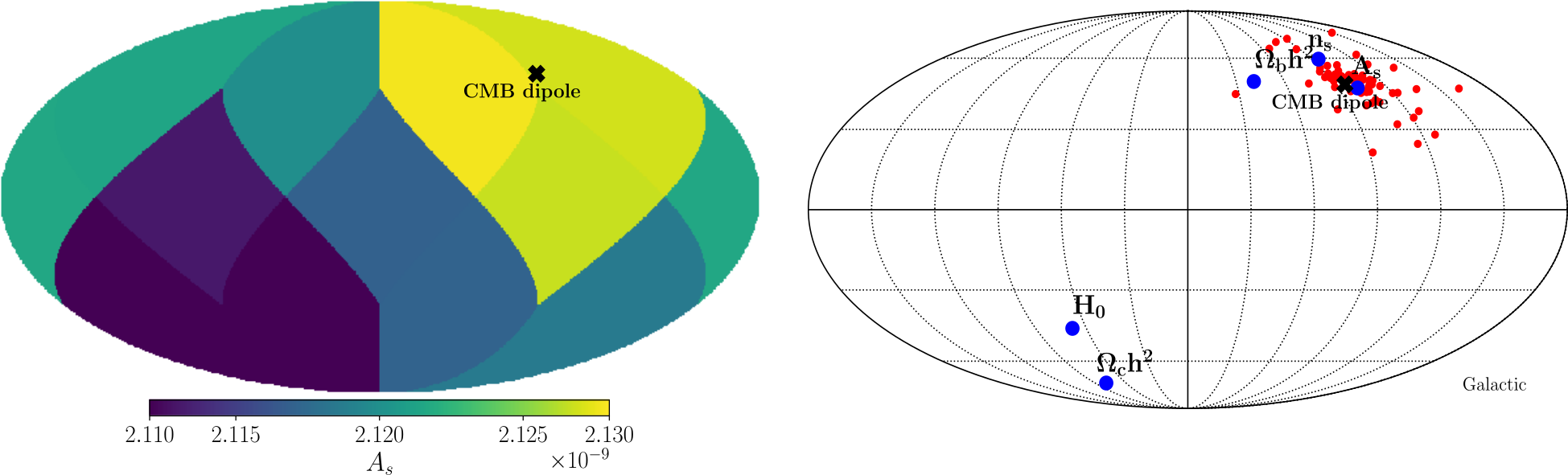}
    \caption{\textit{Left panel}: Mean field for the $A_{s}$ cosmological parameter, obtained by averaging the results from the 600 E2E PR4 simulations for each of the 12 patches. \textit{Right panel}: Directions of the dipoles fitted on the mean field for each of the five cosmological parameters (blue dots). The dipole directions for the mean fields of the $TT$ power spectrum bandpowers are also shown (red dots). In both panels the black cross corresponds to the CMB dipole direction.}
    \label{fig12}
\end{figure}

An intriguing aspect of the analysis is how Doppler boosting affects the cosmological parameters, particularly $A_{s}$. Before fitting a dipole to the parameter maps, we calculate the mean field and standard deviation for each parameter using all simulations, following the same procedure used for the power spectra in the angular clustering analysis. Notably, the mean field for $A_{s}$ exhibits a dipolar feature aligned with the CMB dipole, as shown in the left panel of figure~\ref{fig12}. We then fit dipoles to the mean fields. The right panel of Figure~\ref{fig12} displays the fitted dipoles along with the directions of the dipoles obtained from the mean fields of each bandpower of the power spectrum (red dots). For reference, we also include the direction of the CMB dipole from \cite{2020A&A...643A..42P}. A clear alignment is observed between the $A_{s}$ mean field dipole, the CMB dipole, and the dipoles of the bandpower mean fields. However, this effect is mitigated by subtracting the mean fields prior to the fit. As shown later, the directions in the data are not aligned with the CMB dipole; in particular, the direction for $A_{s}$ is nearly 55 degrees away. This indicates that Doppler boosting does not influence our estimator. Furthermore, after debiasing, the mean field for $A_{s}$ no longer displays a dipolar pattern, suggesting that Doppler boosting is a contributing factor to the observed bias.

\begin{figure}[t!]
    \centering
    \includegraphics[scale = 0.4]{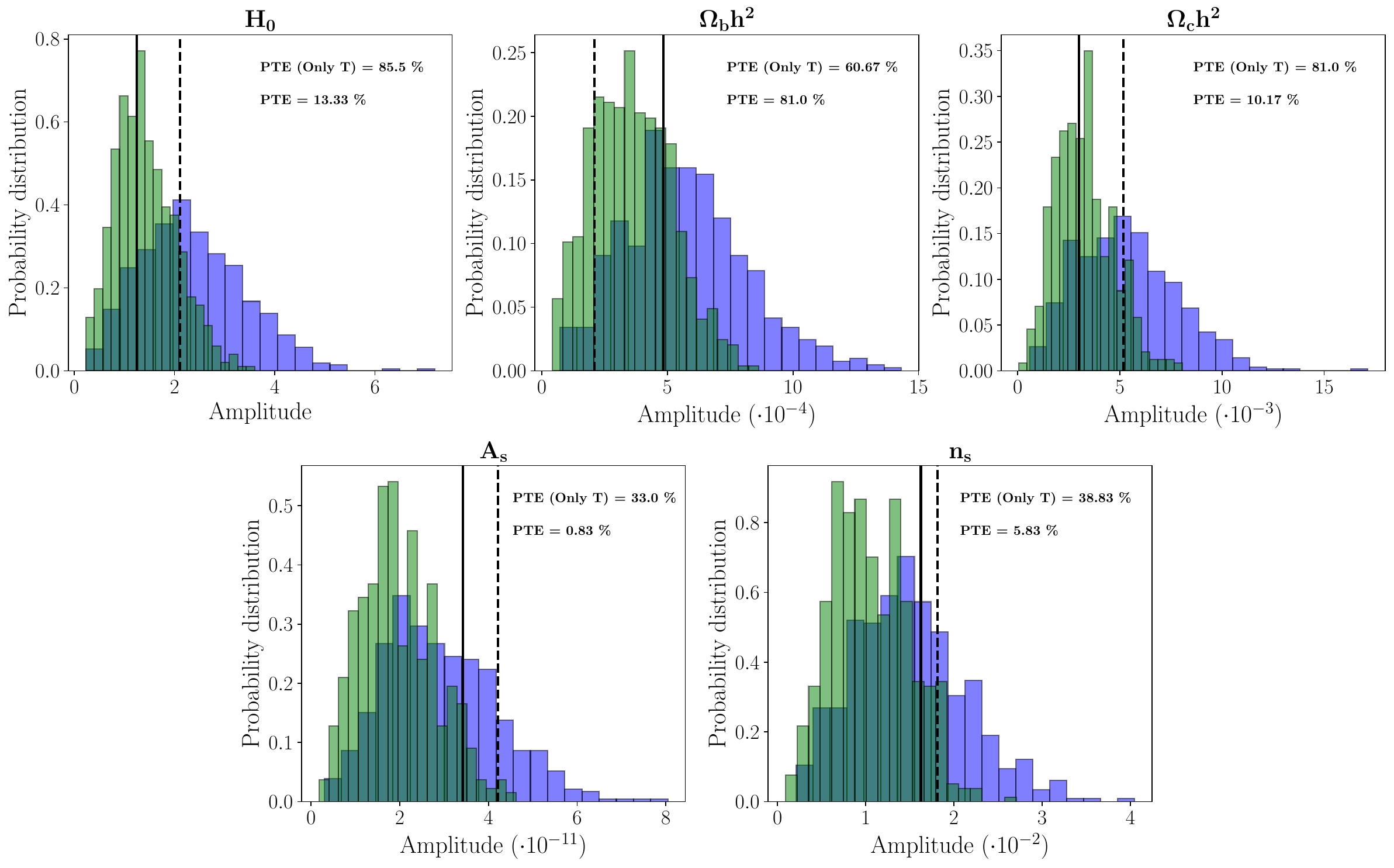}
    \caption{Probability distribution of dipole amplitudes for the 600 E2E PR4 simulations for the five cosmological parameters. Blue distribution represent temperature-only case, while green distribution includes also the polarization information. The black lines correspond to the values observed in the data. Solid line for temperature-only case and dash line for temperature plus polarization. In each case the probability to exceed is also provided.}
    \label{fig16}
\end{figure}

Figure~\ref{fig16} presents the distribution of dipolar amplitudes for the five cosmological parameters. The green distribution is obtained including polarization, while the blue distribution is found using only temperature information. The black vertical lines represent the data, the solid line for the temperature plus polarization case, and dashed line for temperature only. It also includes the probability-to-exceed (PTE) values, representing the percentage of simulations with an amplitude equal to or larger than that observed in the data. In particular, for the temperature plus polarization case, the amplitudes for $H_{0}$, $\Omega_{b}h^{2}$, $\Omega_{c}h^{2}$, and $n_{s}$ are fully consistent with the $\Lambda$CDM predictions. However, there are only 5 simulations from 600 with a larger amplitude for the $A_{s}$ dipole than observed in the data. For the temperature-only analysis, none of the five parameters exhibits an anomalous amplitude. Two effects contribute to this difference. First, the distribution of amplitudes is broader for the temperature-only scenario, consistent with the increased uncertainty in the parameters. Second, the amplitude observed in the data is slightly larger, a trend common to all parameters except $\Omega_{b}h^{2}$.

\begin{figure}[t!]
    \centering
    \includegraphics[scale = 0.8]{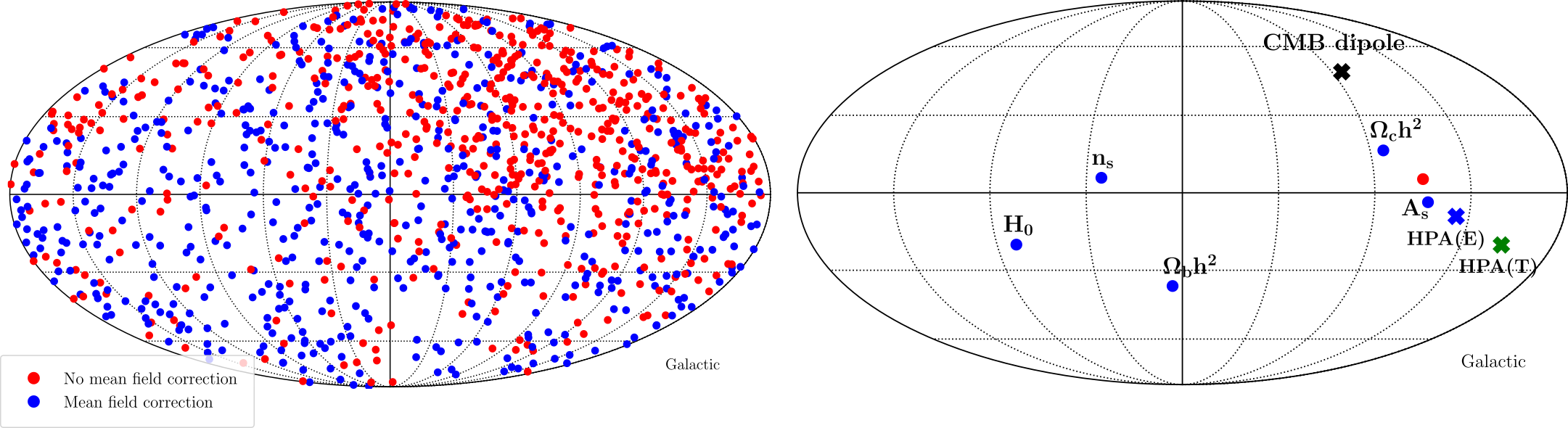}
    \caption{\textit{Left panel}: Dipole directions of the $A_{s}$ parameter for the 600 E2E PR4 simulations before (red dots) and after (blue dots) mean field subtraction. \textit{Right panel}: Dipole directions of the five cosmological parameters observed in the data (blue dots). The dipole direction for $A_{s}$ before mean field subtraction is also shown (red dot). The black cross corresponds to the CMB dipole direction, and the blue and green crosses represent the HPA directions measured in $T$ and polarization $E$-modes, respectively.}
    \label{fig14}
\end{figure}

The right panel in figure~\ref{fig14} shows the dipole directions in the data for all the cosmological parameters. In particular, $A_{s}$, which is the only one showing an anomalous amplitude, is closest to the HPA. The HPA directions used in this analysis are those reported in ref.~\cite{2023JCAP...12..029G}. For reference, we present both HPA directions: one inferred from temperature data alone and the other derived from polarization $E$-modes alone. We also show the direction for the $A_{s}$ dipole without the mean field correction. Even if this is displaced slightly towards the CMB dipole direction, it is not significantly affected by the Doppler boosting effect, contrary to what happens with the simulations (left panel). Moreover, the PTE is still below 1\%. This means that the dipole exhibited by the $A_{s}$ parameter is stronger than the one produced by the Doppler boosting.   

\begin{figure}[t!]
    \centering
    \includegraphics[scale = 0.7]{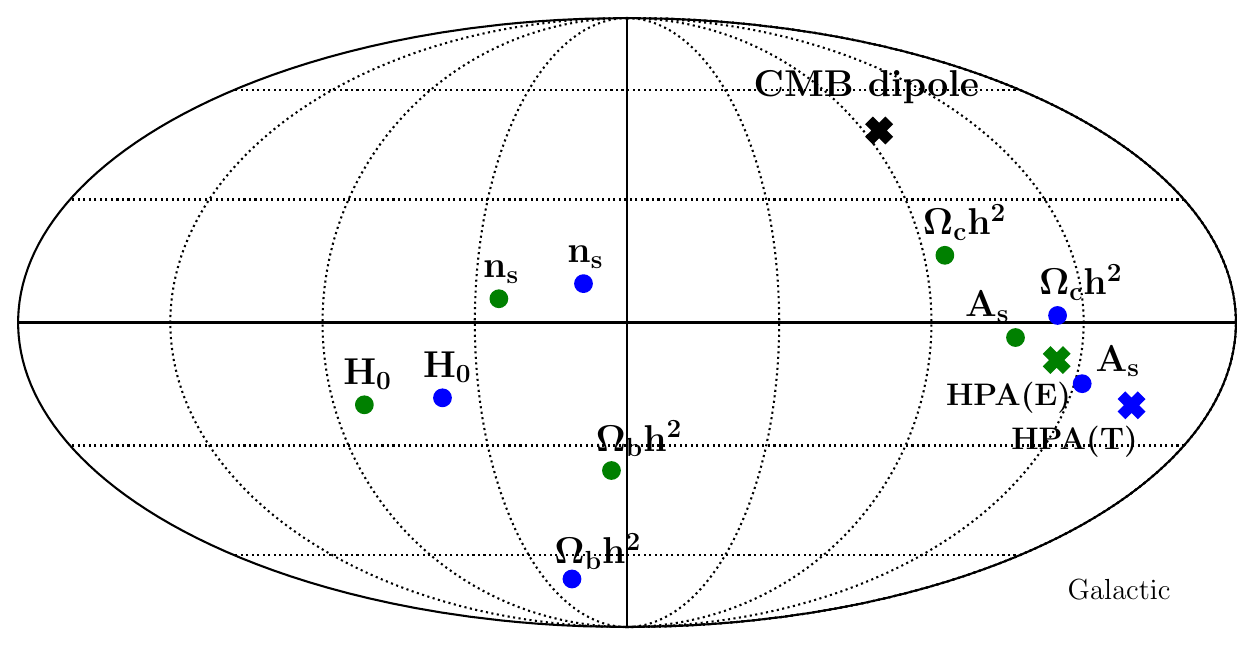}
    \caption{Dipole directions of the five cosmological parameters observed in the data. Green dots correspond to temperature plus polarization case, while blue dots are from temperature-only scenario. The black cross corresponds to the CMB dipole direction, and the blue and red cross represent the HPA directions measured in $T$ and polarization $E$-modes, respectively.}
    \label{fig17}
\end{figure}
Figure \ref{fig17} shows the directions in data for all the cosmological parameters for both scenarios, temperature only (red dots) and including polarization (blue dots). An interesting fact is that even if the amplitude of $A_{s}$ is no longer anomalous, the direction is still close to the HPA direction. In particular, the distance between the $A_{s}$ dipole direction observed in the temperature-only fits and the HPA for temperature data alone is quite similar to the distance between the $A_{s}$ dipole inferred including polarization and the HPA for the polarization $E$-mode signal alone. Additionally, we see that the directions for temperature only are not far away from the directions obtained including polarization. 

It is well-established that $A_{s}$ and $\tau$ are degenerate, and the uncertainty in $A_{s}$ increases when $\tau$ is not fixed. As a result, the analysis presented here examines the conditioned distribution for $A_{s}$, rather than the distribution marginalized over the $\tau$ value. However, we are unable to quantify by how much our results, particularly the PTE for $A_{s}$, might change if $\tau$ is not fixed. The primary limitation is that we cannot extract meaningful information about $\tau$ from such small sky patches. 

To address this, we performed a set of tests. First, we run the pipeline with $\tau$ as a free parameter, imposing a reasonable bound in \texttt{iMinuit} between 0.03 and 0.09 — ten times the Planck error bar on $\tau$. In almost all simulations, the $\tau$ values clustered near the boundaries of the allowed range, reflecting that no information can be determined about $\tau$ alone. However, the distribution of $A_{s}e^{-2\tau}$ seems to be similar to that obtained by fixing $\tau$. In other words, the data can constrain the combination fairly well regardless of whether $\tau$ is fixed. We also run this case for the data, finding similar behavior. The final $p$-value for the combination $A_{s}e^{-2\tau}$ is at the level of 2\%. Therefore, it seems that there is an anomaly that is best captured by the combination $A_{s}e^{-2\tau}$, but once $\tau$ is fixed, it is propagated to $A_{s}$, because both quantities are related by a constant.

Additionally, we run the pipeline with a Gaussian prior on $\tau$. Unlike the Bayesian approach, the minimizer can not deal with the prior and consistently converged to the maximum of it, yielding results essentially identical to those obtained with fixed $\tau$. Finally, we consider the case in which the first bin, covering multipoles between 2 and 31, is included in the likelihood. Although this bin carries information about $\tau$, the inferred values are still clustered near the bounds. This is expected, given the small $f_{\mathrm{sky}}$, which produces large error bars. Moreover, on these scales, the Gaussian approximation on the likelihood is no longer a good approximation, even if the bin size makes it more Gaussian following the central limit theorem. For all of these reasons, we decided to focus on the case where $\tau$ is fixed.

\section{Conclusions} \label{Con}

In this study, we have performed two complementary analyses of the Planck PR4 detector splits, which were cleaned using the \sevem\ component separation method, in order to test the statistical isotropy of the Universe. Our approach employed well established techniques for both power spectrum estimation (\texttt{MASTER}) and cosmological parameter fitting (\texttt{iMinuit}). In both analyses, we used the $TT$, $TE$, and $EE$ binned power spectra computed over 12 independent sky regions defined by the $N_{\mathrm{nside}}$ = 1 \texttt{HEALPix} scheme. These patches overlap with the Planck confidence mask, so the effective $f_{\mathrm{sky}}$ for each region varies between 2\% and 8\%.

The first analysis focuses on the angular clustering feature. Previous Planck releases reported an unexpected alignment in directions derived from temperature power distribution maps over a wide range of angular scales. We confirm this alignment over a broader multipole range. The $p$-value -- defined as the fraction of simulations exhibiting a Rayleigh statistic greater than that observed in the data -- is below 1\% for multipoles between $\ell$ = 200 and 2000. We suggest that the discrepancy with earlier results for multipoles above $\ell$ = 1000 may arise from artificial clustering due to the absence of apodization, which can introduce correlations between the final bins. Additionally, the hint of an anomaly in the $E$ modes that was present in PR3 disappears, with only a couple of bins showing $p$-values below 1\%. In particular, this occurs in the same multipole range where $TT$ and the alignment between $TT$ and $EE$ begins to exhibit anomalous behavior. However, the interpretation remains unclear, as removing the first bin or considering only the cosines between dipoles within the same bin reduces the statistical significance. Given that the anomalous behavior in the $E$ modes takes place within a narrow multipole range, it is most likely a consequence of the look elsewhere effect. A similar reasoning can be applied to a few bins close to 100\% for $TE$.

In our second analysis, we examine the potential presence of dipolar variations in the cosmological parameters. Using \texttt{iMinuit}, we maximized the likelihood in each of the 12 independent patches using the previously computed $TT$, $TE$, and $EE$ binned power spectra. Given the limited sky fraction available in each of these patches, we can not access the large scale $E$ modes and $\tau$ needs to be fixed. Fitting a dipole on the resulting maps reveals that all parameters, except for $A_{s}$, are consistent with the standard cosmological model. A hint of anomalous behaviour was detected in $A_{s}$, with only 5 out of 600 simulations showing an amplitude as extreme as that observed in the data. This anomaly appears to be associated with the combination of $A_{s}$ and $\tau$, $A_{s}e^{-2\tau}$, as it is the quantity best constrained by the data. Once $\tau$ is fixed, the anomaly propagates to $A_{s}$. The direction of the dipole is close to that of the hemispherical power asymmetry, and it is also located in the region where the intensity power spectrum bandpower dipoles are clustered, suggesting a potential relation between these anomalies. Although the anomaly remains robust to variations in the choice of $\ell_{min}$ and $\ell_{max}$
(see Appendix \ref{Appendix_A}), it disappears when the $TE$ and $EE$ spectra are excluded.

An additional outcome validating our pipeline is the successful detection of the Doppler boosting effect in simulations, which is observed at both the power spectrum level and in the $A_{s}$ parameter. This effect can be effectively encoded and subtracted in the mean field prior to dipole fitting, and it is unlikely to be the origin of the anomaly given that the direction observed in $A_{s}$ is approximately 55 degrees away from the CMB dipole direction.

Our results do not agree with some previous analyses which claim a strong evidence for the violation of the cosmological principle of isotropy indicated by cosmological parameter variations across the CMB sky. This may be attributed to methodological differences. In particular, our analysis uses completely independent sky regions and relies on covariances fully derived from the end-to-end simulations provided by the Planck team. Finally, the evidence for an anomaly in $A_{s}e^{-2\tau}$ combination is modest and not entirely conclusive. Future analyses with improved polarization data will be crucial to further clarify these findings.

\appendix
\section{Robustness of Results} \label{Appendix_A}

In order to test the robustness of the results on the anomalous amplitude of $A_{s}$, we consider a set of pipeline runs while modifying some parameters. We run the following cases:

\begin{itemize}
    \item \textbf{Debiased case ($TTTEEE$): } Same pipeline as in the main analysis but subtracting the unknown transfer function from $TT$, $TE$, and $EE$. Remember that these transfer functions are calibrated with the simulations. The main goal of this run is to validate that the transfer function is not introducing any anisotropy in the cosmological parameters.
    \item \textbf{Debiased case (only $T$): } Same pipeline but only for the temperature data. 
    \item \textbf{Different $\boldsymbol{\tau}$: } In this case we modify the value of $\tau$. We fix $\tau$ to 0.0544,  which was the best-fit value inferred from Planck Data Release 3 \cite{2020A&A...641A...6P}, instead of the input $\tau$ value for the E2E simulations. In this sense, we can check if the anomaly persists when the fixed value is not the correct one, which may be the case for the data.
    \item \textbf{$\boldsymbol{\ell_{\mathrm{max}}}$ = 1000: } In this case we remove some multipole bins for $TE$ and $EE$. In particular, we consider $\ell_{\mathrm{max}}^{TE} \sim $ 1500 and $\ell_{\mathrm{max}}^{EE} \sim $ 1000.
    \item \textbf{$\boldsymbol{\ell_{\mathrm{min}}}$ = 62: } In this final case, we remove the large angular scale bins, and start at $\ell$ = 62 for $TT$, $TE$, and $EE$.  
\end{itemize}

The values of the PTEs for the five cosmological parameters are shown in table~\ref{tab_appA}, together with the values obtained in the main analysis.
Notably, both the main and the debiased cases produce very similar results, finding again that the PTE for $A_{s}$ is below 1\% if polarization is included. The results also remain very stable when using an alternative value of $\tau$.
Furthermore, this result appears to be robust against different choices of multipole cuts. In particular, if the last 500 multipoles of the $E$ modes are removed, and only the first 1000 multipoles are considered, the PTE remains below 2\%. On the other hand, if the first bin, which contains multipoles between $\ell$ = 32 and $\ell$ = 61, is removed, the PTE increases to 2.7\%. 
Although this decrease in significance may be attributed to the loss of statistical information when excluding some multipoles, it may also suggest that the anomaly has a stronger contribution from large scales. Notably, it is precisely at large scales where the known CMB anomalies are found in the temperature data, particularly the HPA.

\begin{table}[t!]
\centering
\begin{tabular}{c|ccccc}
\hline
\textbf{Scenario}                                  & \textbf{PTE ($\boldsymbol{H_{0}}$)} & \multicolumn{1}{l}{\textbf{PTE ($\boldsymbol{\Omega_{b}h^{2}}$)}} & \multicolumn{1}{l}{\textbf{PTE ($\boldsymbol{\Omega_{c}h^{2}}$)}} & \multicolumn{1}{l}{\textbf{PTE ($\boldsymbol{A_{s}}$)}} & \multicolumn{1}{l}{\textbf{PTE ($\boldsymbol{n_{s}}$)}} \\ \hline
\textbf{Main (Inc. Pol)}                           & 13.3\%                             & 81.0\%                                                              & 10.2\%                                                           & 0.83\%                          & 5.8\%                                                  \\
\textbf{Main (Only T)}                             & 85.5\%                              & 60.7\%                                                           & 81.0\%                                                              & 33.0\%                                                    & 38.8\%                                                 \\
\textbf{Debiased (Inc. Pol)}                         & 14.3\%                             & 81.3\%                                                           & 10.8\%                                                           &  0.83\%   & 6.0\%                                                     \\
\textbf{Debiased (Only T)}                           & 86.8\%                             & 62.7\%                                                           & 81.7\%                                                           & 33.5\%                                                  & 38.5\%                                                  \\
\textbf{$\boldsymbol{\tau}$ = 0.0544}              & 13.5\%                              & 80.7\%                                                           & 10.3\%                                                           & 0.83\%                          & 5.8\%                                                  \\
\textbf{$\boldsymbol{\ell_{\mathrm{max}}}$ = 1000} & 21.2\%                             & 86.3\%                                                           & 11.7\%                                                           &  1.5\%    & 10.2\%                                                 \\
\textbf{$\boldsymbol{\ell_{\mathrm{min}}}$ = 62}   & 22.0\%                                & 79.2\%                                                           & 18.8\%                                                           & 2.7\%                          & 12.8\%                                                
\end{tabular}
\caption{PTEs of the five cosmological parameters for seven different scenarios as a test of robustness.}
\label{tab_appA}
\end{table}

We pay particular attention to the case where $\tau$ is fixed to a different value. In this scenario, the other cosmological parameters remain largely unaffected except for $A_s$, which is expected given their correlation — a smaller $\tau$ produces a smaller $A_s$, and vice versa. However, because $\tau$ is fixed to the same value in all patches, $A_s$ is shifted uniformly, leaving the dipole amplitude unchanged. In other words, fixing $\tau$ to the same values for all patches allows the same amount of fluctuations in $A_{s}$ independently of the fixed value. Consequently, we obtain a similar PTE.

\begin{figure}[t!]
    \centering
    \includegraphics[scale = 0.5]{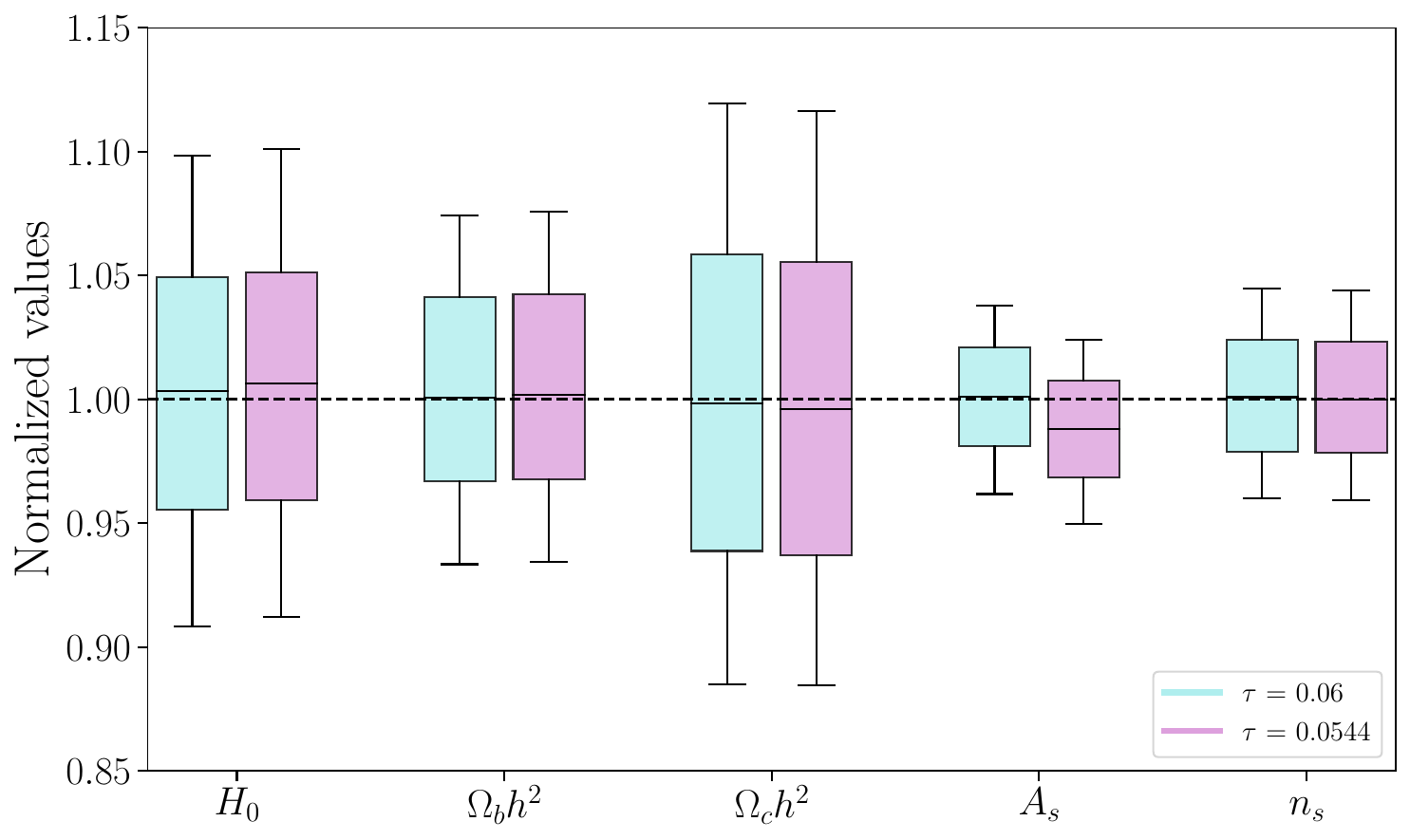}
    \caption{Distribution of the five cosmological parameters for the 600 E2E simulations computed on patch number 4 ($f_{\mathrm{sky}} \approx $ 2\%) and fixing $\tau$ to two different values. Distributions are normalized to the input values. The boxes represent 68\% of the probability, while the whiskers include 95\%.}
    \label{fig18}
\end{figure}

\begin{figure}[t!]
    \centering
    \includegraphics[scale = 0.4]{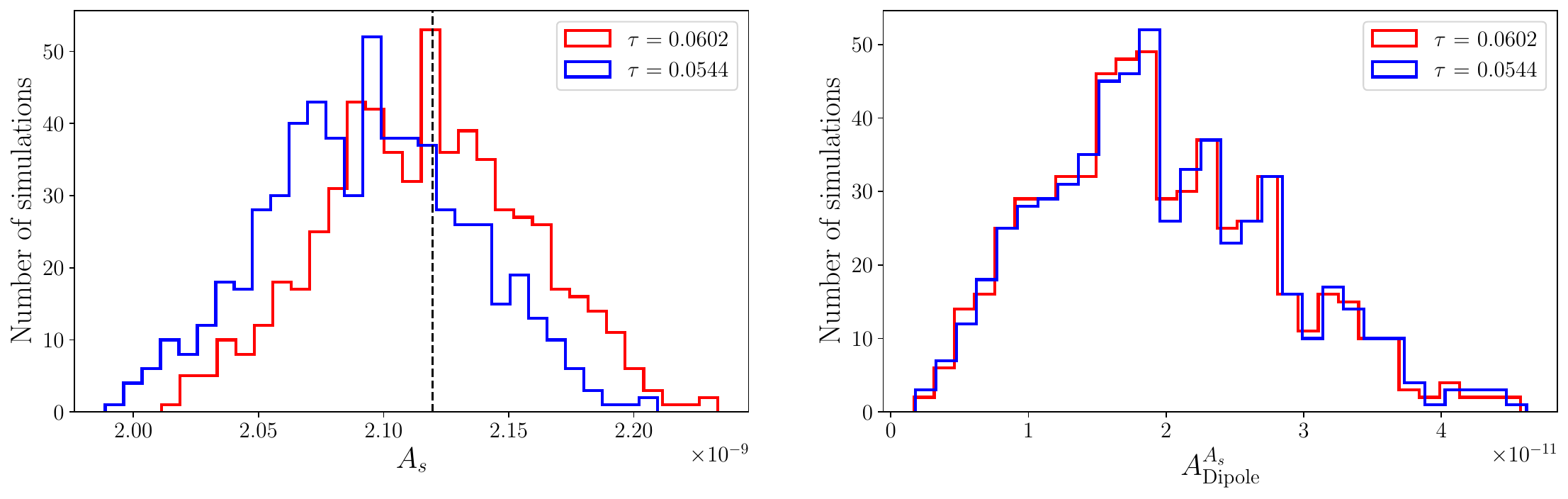}
    \caption{\textit{Left panel: } Distribution of the $A_{s}$ parameter for the 600 E2E simulations fixing $\tau$ to 0.0602 (in red) and 0.0544 (in blue). The black dashed line corresponds to the input $A_{s}$ value for the simulations. \textit{Right panel: } Distribution of the dipole amplitudes of $A_{s}$ for the 600 E2E PR4 simulations. Again the red colour is for $\tau$ = 0.0602, while blue is used for $\tau$ = 0.0544.}
    \label{fig19}
\end{figure}

Figure~\ref{fig18} shows results for the five cosmological parameters with two fixed values of $\tau$, computed for patch number 4. It is clear that, except for $A_{s}$, the other parameters remain unaffected, while $A_{s}$ is reduced. However, the input value is still within 1$\sigma$. Similarly, the left panel of figure~\ref{fig19} shows the shift in $A_{s}$. As previously mentioned, this shift does not affect the dipole amplitudes, as shown in the right panel. 

\acknowledgments
 
We thank Patricio Vielva and Reijo Keskitalo for valuable advice and comments. CGA, EMG and RBB would like to thank the Spanish MCIN/AEI/10.13039/501100011033, project refs. PID2019-110610RB-C21 and PID2022-139223OB-C21 (the latter funded also by European Union NextGenerationEU/PRTR) for finantial support, and the support from the Universidad de Cantabria and Consejer\'{\i}a de Educaci\'on, Formaci\'on Profesional y Universidades  from Gobierno de Cantabria via the \emph{Actividad estructural para el desarrollo de la investigaci\'on del Instituto de F{\'\i}sica de Cantabria} project.
CGA also thanks the funding from the Formaci\'on de Personal Investigador (FPI, Ref. PRE2020-096429) program of the Spanish Ministerio de Ciencia, Innovaci\'on y Universidades. 
We acknowledge Santander Supercomputacion support group at the University of Cantabria who provided access to the supercomputer Altamira Supercomputer at the Institute of Physics of Cantabria (IFCA-CSIC), member of the Spanish Supercomputing Network, for performing simulations/analyses. FKH acknowledges the use of computing resources provided by UNINETT Sigma2 - the National Infrastructure for High Performance Computing and  Data Storage in Norway. The presented results are based on observations obtained with Planck\footnote{\url{http://www.esa.int/Planck}}, an ESA
science mission with instruments and contributions directly funded
by ESA Member States, NASA, and Canada. The results of this paper have been derived using the \texttt{HEALPix} package \cite{Gorski:2004by}, and the \texttt{healpy} \cite{Zonca2019}, \texttt{numpy} \cite{2020Natur.585..357H}, \texttt{matplotlib} \cite{2007CSE.....9...90H} and \texttt{scipy} \cite{2020NatMe..17..261V} \texttt{Python} packages.




\bibliographystyle{JHEP}
\bibliography{Bibliography}

\end{document}